\pdfoutput=1
\documentclass[referee]{raa}

\usepackage{graphicx,times}
\usepackage[a4paper]{geometry}
\usepackage{natbib}
\usepackage{amssymb,amsmath}
\usepackage{subcaption}
\usepackage{array}
\usepackage{xcolor}
\usepackage{tabularx}
\usepackage{float}
\bibpunct{(}{)}{;}{a}{}{,}

\usepackage[pagebackref=true]{hyperref}

\begin{document}

   \title{Derivative Spectroscopy and its application at detecting the weak emission/absorption lines
   }

 \volnopage{ {\bf 20XX} Vol.\ {\bf X} No. {\bf XX}, 000--000}
   \setcounter{page}{1}

   \author{Lihuan Yu, Jiangdan Li, Jinliang Wang, Jiajia Li, Jiao Li, Qiang Xi, Zhanwen Han*
   }

   \institute{
    Yunnan Observatories, Chinese Academy of Sciences (CAS), 396 Yangfangwang, Guandu District, Kunming 650011, P. R. China;{\it Yulihuan@ynao.ac.cn; zhanwenhan@ynao.ac.cn}
    \and
        Key Lab of Space Astronomy and Technology, National Astronomical Observatories, Chinese Academy of Sciences, Beijing 100101, P. R. China
             \\
\vs \no
   {\small Received 20XX Month Day; accepted 20XX Month Day}
}

\abstract{The development of spectroscopic survey telescopes (SSTs) like Large Sky Area Multi-Object Fiber Spectroscopic Telescope (LAMOST), Apache Point Observatory Galactic Evolution Experiment (APOGEE), and Sloan Digital Sky Survey (SDSS) has opened up unprecedented opportunities for stellar classification. Specific types of stars, such as early-type emission-line stars and those with stellar winds, can be distinguished by the profiles of their spectral lines. In this paper, we introduce a method based on derivative spectroscopy(DS) designed to detect signals within complex backgrounds and provide a preliminary estimation of curve profiles. This method exhibits a unique advantage in identifying weak signals and unusual spectral line profiles when compared to other popular line detection methods. We validated our approach using synthesis spectra, demonstrating that DS can detect emission signals three times fainter than Gaussian fitting. Furthermore, we applied our method to 579,680 co-added spectra from LAMOST Medium-Resolution Spectroscopic Survey(LAMOST-MRS), identifying 16,629 spectra with emission peaks around the $\rm{H\alpha}$ line from 10,963 stars. These spectra were classified into three distinct morphological groups, resulting in nine subclasses as follows. 1. Emission peak above the pseudo-continuum line (single peak, double peaks, emission peak situated within an absorption line, P Cygni profile, Inverse P Cygni profile) 2. Emission peak below the pseudo-continuum line (sharp emission peak, double absorption peaks, emission peak shifted to one side of the absorption line) 3. Emission peak between the pseudo-continuum line.
\keywords{methods: data analysis - line: identification - line: profiles - techniques: spectroscopic - - techniques: radial velocities}
}

   \authorrunning{L.-H. Yu et al. }            
   \titlerunning{Derivative Spectroscopy}  
   \maketitle

%
\section{Introduction}           
\label{sect:intro}

Selecting the desired stellar spectra from a massive dataset is a key challenge. The continuum, spectral lines, and profile features of these lines represent a portion of stellar characteristics. In this paper, our focus lies on spectral line shapes, which can be broadly classified into two groups: those devoid of emission lines and those exhibiting distinct emission-line features. The latter group encompasses various subclasses:
\\

1. Single-peak and Double-peak types, distinguished by the number and wavelengths of emission line
\newline

\vspace{-20pt}
\hspace{5pt} peaks.\citep[e.g.][]{Zhang+2022}

2. Emission within absorption lines, P Cygni, and Inverse P Cygni profiles, which depend on the relative
\newline

\vspace{-20pt}
\hspace{5pt}
positions of emission and absorption lines.\citep[e.g.][]{Snow+1994}

3. Emission blend or sharp emission line profiles, classified based on the number and characteristics of
\newline

\vspace{-20pt}
\hspace{5pt}
the emission features.\citep{Traven+2015}
\\

Various types of $\rm{H\alpha}$ lines serve as valuable references for classifying stellar systems. For instance, double-peak emission lines can potentially arise from processes such as jet emission from the stellar polar regions, which may exhibit a significant inclination with the observer's line of sight. Additionally, the presence of accretion disks could also lead to the emergence of such spectral features\citep[e.g.][]{Bromley+1997}. Double-peak absorption lines, on the other hand, may be indicative of double-lined spectroscopic binary systems (SB2), where the two absorption peaks could be attributed to a binary system\citep{Aoki+2014}. Stars with stellar winds or mass accretion can be identified using P Cygni or Inverse P Cygni profiles, allowing for the calculation of wind velocities based on the wavelength of the absorption peaks.

General methods for classifying $\rm{H\alpha}$ lines include Gaussian fitting, machine learning, and analyzing spectral line asymmetry, among others.\citep[e.g.][]{Traven+2015}

The fitting technique, which includes methods like least squares\citep{Merriman+1877}, maximum likelihood\citep{Rossi+2018}, and affine invariant Markov chain Monte Carlo (MCMC)\citep{Foreman+2013}, is a general approach for separating different components of a spectrum. This method works exceptionally well when we have a prior understanding of the spectral shape. Unfortunately, it encounters challenges when dealing with spectra that fall outside the adjustable parameter space of the prior predictive fitting method. This is a common issue when dealing with unusual spectra among a large sample.

Machine learning techniques, including K-Nearest Neighbor (KNN)\citep{Beyer+1999}, Random Forest (RF)\citep{Breiman+2001}, AdaBoost\citep{Hastie+2009}, Naive Bayes\citep{Webb+2010}, logistic regression\citep{Wright+1995}, Support Vector Machine (SVM)\citep{Noble+2006}, and Artificial Neural Network\citep{Jain+1996}, have been utilized for the classification of spectra. These techniques rely on specific characteristics of spectral lines for classification, as described in \cite{Zhang+2022}. The process involves creating a training dataset by selecting a subset of features extracted from spectra, such as equivalent width or full width at half maximum. The trained model is subsequently used to classify spectra based on these extracted features.

It is intuitive to speculate that the detection results of machine learning models depend on the training dataset we use. This dataset should ideally consist of samples that we already have or samples separated from SST (spectroscopic survey telescope) data. However, it is challenging to create a comprehensive training dataset for rare spectral lines with complex structures, such as the P Cygni or Inverse P Cygni profiles. These lines are difficult for researchers to separate or identify from thousands of spectral data, making it challenging to build a complete training dataset.

\begin{figure}
    \centering
    \includegraphics[width=10.0cm, angle=0]{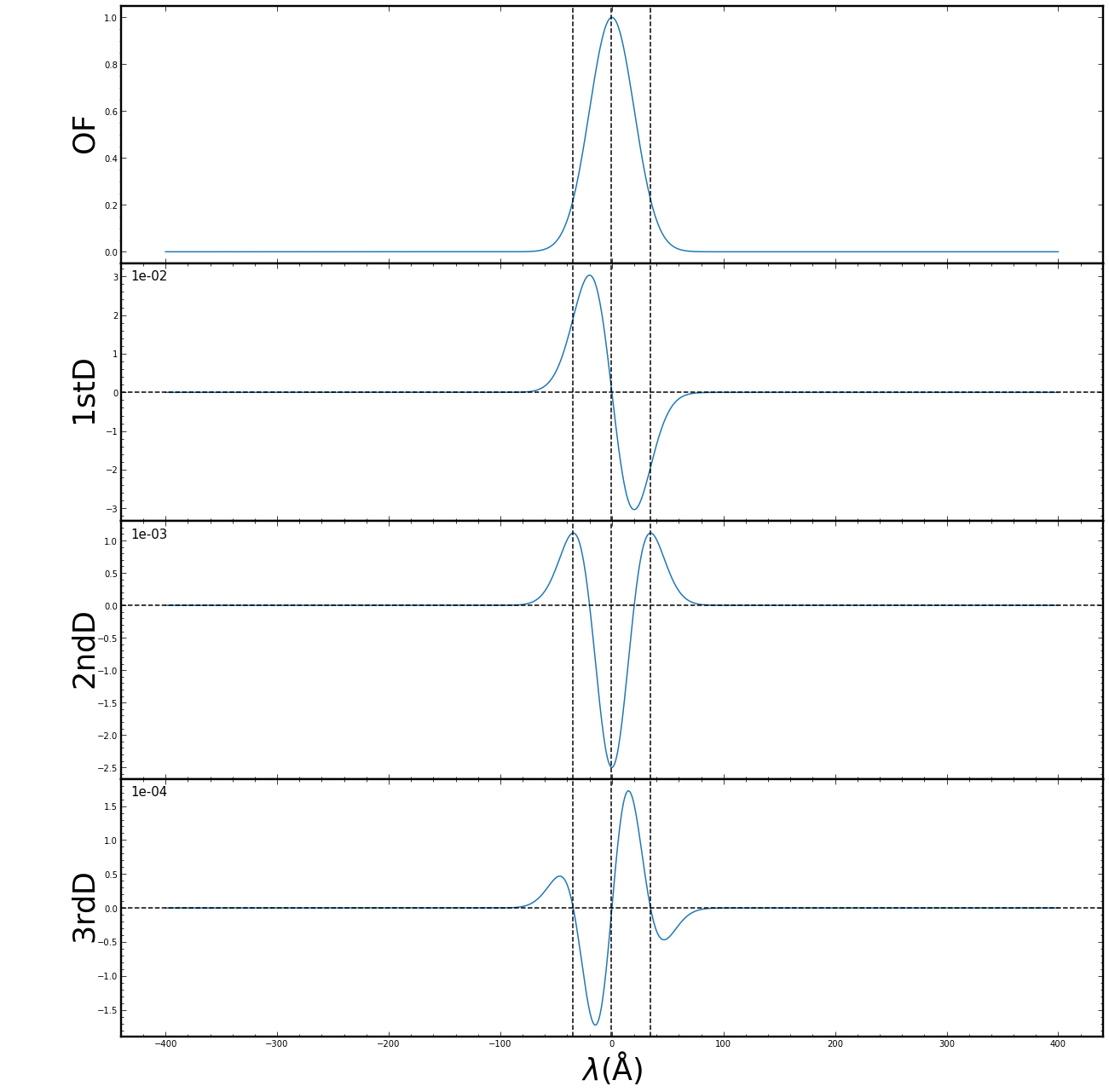}

    \caption{Derivative spectra of a Gaussian function. The original flux(OF) in the first panel represents a Gaussian function, and its first derivative(1st D), second derivative(2nd D), and third derivative(3rd D) are displayed in the three panels below. The three vertical black dashed lines correspond to the three zero points of the third derivative, where the zero point in the rising part corresponds to the peak of the original spectrum.}
    \label{Fig1}
\end{figure}

In addressing the limitations of the above methods, we turned to the derivative technique, which was employed in analytical spectrophotometry by chemists during the 1980s to detect and pinpoint the wavelengths of complex spectrum signals that were challenging to resolve \citep{Haver+1982}. The general calculation of a derivative involves dividing the difference between the original spectrum $f(\lambda)$ and the same spectrum displaced by a finite wavelength $f(\lambda + \rm{\Delta}\lambda)$ by that finite wavelength $\rm{\Delta} \lambda$, which is associated with the midpoint of the finite wavelength, resulting in $\rm{\frac{d\it{f}}{d\it{\lambda}}}(\lambda + \rm{\frac{1}{2}}\rm{\Delta}\lambda)$ and the higher derivatives are obtained by repeating this procedure the desired number of times \citep{Warren+1979}. In Figure \ref{Fig1}, the original curve created by a Gaussian function provides three derivatives - the first derivative, the second derivative, and the third derivative - beneath it through this process. The positive part of the third derivative exhibits a maximum value, two zeros, and a full amplitude (zero - maximum - zero), which correspond to the rising part of the original curve, while the negative part of the third derivative arises from the declining part of the curve, as expected.

This paper introduces DS, a method aimed at extracting spectral line features and classifying profile types. In Section \ref{sec:Metho}, we provide a detailed explanation of our methodology and assess its effectiveness using synthesized spectra. Moving on to Section \ref{sec:tes}, we apply DS to LAMOST-MRS data, leading to the identification of nine distinct subclasses. Subsequently, in Section \ref{sect:Dis}, we engage in a comparative discussion involving DS, machine learning, and Gaussian fitting. Finally, our conclusions are presented in Section \ref{sec:con}.

\section{Method and Derivative Spectroscopy(DS)}
\label{sec:Metho}

\subsection{Basic theory}

Derivative spectroscopy offers a technique for enhancing the resolution of spectra and accurately isolating weak signals from background noise, making it possible to detect the maxima of spectral profiles more precisely \citep{Frederic+1968}. The improved spectral resolution is achieved because the width of peaks becomes narrower with higher-order derivatives\citep{Anthony+1983}. In analytical scenarios, the Full Width Half Maximum (FWHM) of the fourth derivative spectrum, which is composed of a Lorentzian function, reduces to one-quarter of the FWHM of the original spectrum. This means that extreme values that might be obscured by overlapping components become more pronounced in higher-order derivative spectra\citep{Warren+1979}.

\begin{figure}
    \centering
    \includegraphics[width=10.0cm, angle=0]{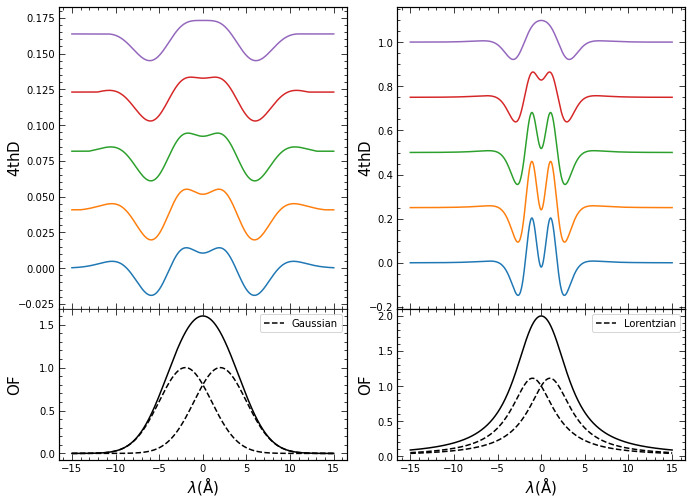}
    \caption{The bottom image on the left illustrates a profile(solid line) formed by the combination of two Gaussian functions(dashed lines) with a 4 Å separation. The fourth derivative(4th D) of this profile is displayed on the upper panel, with the blue curve representing the analytical spectrum. Additionally, there are orange, green, red, and purple curves, which are obtained by computing the differences of the profile with intervals of 2, 4, 6, and 8 Å, respectively. The bottom image on the right depicts the same scenario but with two Lorentzian functions separated by 2 Å. In this case, there are four different spectra lines with intervals of 1, 2, 3, and 4 Å.}
    \label{Fig2}
\end{figure}

In Fig \ref{Fig2}, there are spectra composed of two Gaussian functions (left) and two Lorentzian functions (right), with their fourth derivatives displayed in the top two panels. It's evident that in both cases, the two components with maxima are separated, and the spectrum composed of Lorentzian functions, which have steeper slopes, is divided more significantly.

It is reasonable to use derivative spectroscopy for correcting spectral backgrounds. Every spectral profile can be viewed as a general function that can be expanded into a polynomial form\citep{Pourahmadi+1984}, as follows:

\begin{equation}
    f(\lambda) = f(\lambda_{\rm{0}}) + f'(\lambda_{\rm{0}}) (\lambda - \lambda_{\rm{0}}) + \cdots + \frac{f^{(n)}(\lambda_{\rm{0}})}{n!} x^n\label{equ:1}
\end{equation}

The first term, $f(\lambda_{\rm{0}})$, which contains information about the background of the spectrum, disappears in the first derivative, while the other terms containing information about the slope and structure of the spectrum are retained.

\subsection{Noise and Smooth}
\label{Noi}

However, there is a significant disadvantage to the derivative technique, namely, that the signal-to-noise ratio (SNR) deteriorates as you move to progressively higher derivative orders\citep{Haver+1981}. This was illustrated by \cite{Haver+1981} using an experimental spectrum composed of a series of amplitudes without a signal, taken at discrete and equally spaced wavelength increments ($a_{\rm{1}}$, $a_{\rm{2}}$, $a_{\rm{3}}$, ...). The noise, which is assumed to be independent of the amplitude of the signal and follows a Gaussian amplitude probability distribution, can be expressed as the standard deviation of all the elements in the series. The standard deviation of the nth-order derivative can be calculated using the rules of error propagation:

\begin{equation}
    \sigma_n = \sigma_{\rm{0}} \rm{\sqrt{{\sum\limits_{\it{m}}}(\frac{\it{n}!}{(\it{n} - \it{m})!\it{m}!})}}\label{epu:2}
\end{equation}

In the equation\ref{epu:2}, $n$ represents the order of the derivative, and $\sigma_{\rm{0}}$ is the standard deviation of the original (zeroth order) series. It's evident that the value of $\sigma_n$ increases with the order $n$. To address this issue and improve the SNR of derivative spectra, it's necessary to apply some form of low-pass filtering or smoothing during the differentiation process\citep{Haver+1982}. This smoothing is achieved through the convolution of the data series with a smoothing function composed of a set of weighting coefficients\citep{Wand+1994}. Examples of such smoothing filters include the average filter, median filter, Gaussian filter, Wiener filter, Savitzky-Golay filter, and more\citep{Gonzales+1987,Schafer+2011}.

\subsection{Method}
\label{Med}

To address this limitation, we have considered two methods. The first method involves fitting subsets of the original series surrounding the data point $x_0$ with a polynomial through convolution\citep{Press+1990}:

\begin{equation}
    y = a_{\rm{0}} + a_{\rm{1}} (x - x_{\rm{0}}) + a_{\rm{2}} (x - x_{\rm{0}})^{\rm{2}} + \cdots\cdots
\end{equation}

where $a_{\rm{0}}$ represents the original (zeroth-order) series, $a_{\rm{1}}$ corresponds to the first-order derivative series, $a_{\rm{2}}$ corresponds to the second-order derivative series, and so on.

The second method involves convolving the signal with the derivative of a Gaussian kernel. This process results in the derivative of the original series and is based on the properties of convolution between two generalized functions\citep{Bracewell+1966}.

\begin{equation}
    \frac{\rm{d}}{\rm{d}\it{x}}(\it{f * y}) = \frac{\it{f}}{\rm{d}\it{x}} * \it{g} = \it{f} * \frac{\rm{d}\it{g}}{\rm{d}\it{x}}
\end{equation}

where $\frac{\it{f}}{\rm{d}\it{x}}$ and $\frac{\rm{d}\it{g}}{\rm{d}\it{x}}$ represent the first derivatives of the generalized functions $\it{f}$ and $\it{g}$, respectively. Convolution of the spectrum with the first derivatives of the Gaussian kernel is equivalent to convolving the first derivatives of the spectrum with the Gaussian kernel.

\begin{figure}
    \centering
    \includegraphics[width=15.0cm, angle=0]{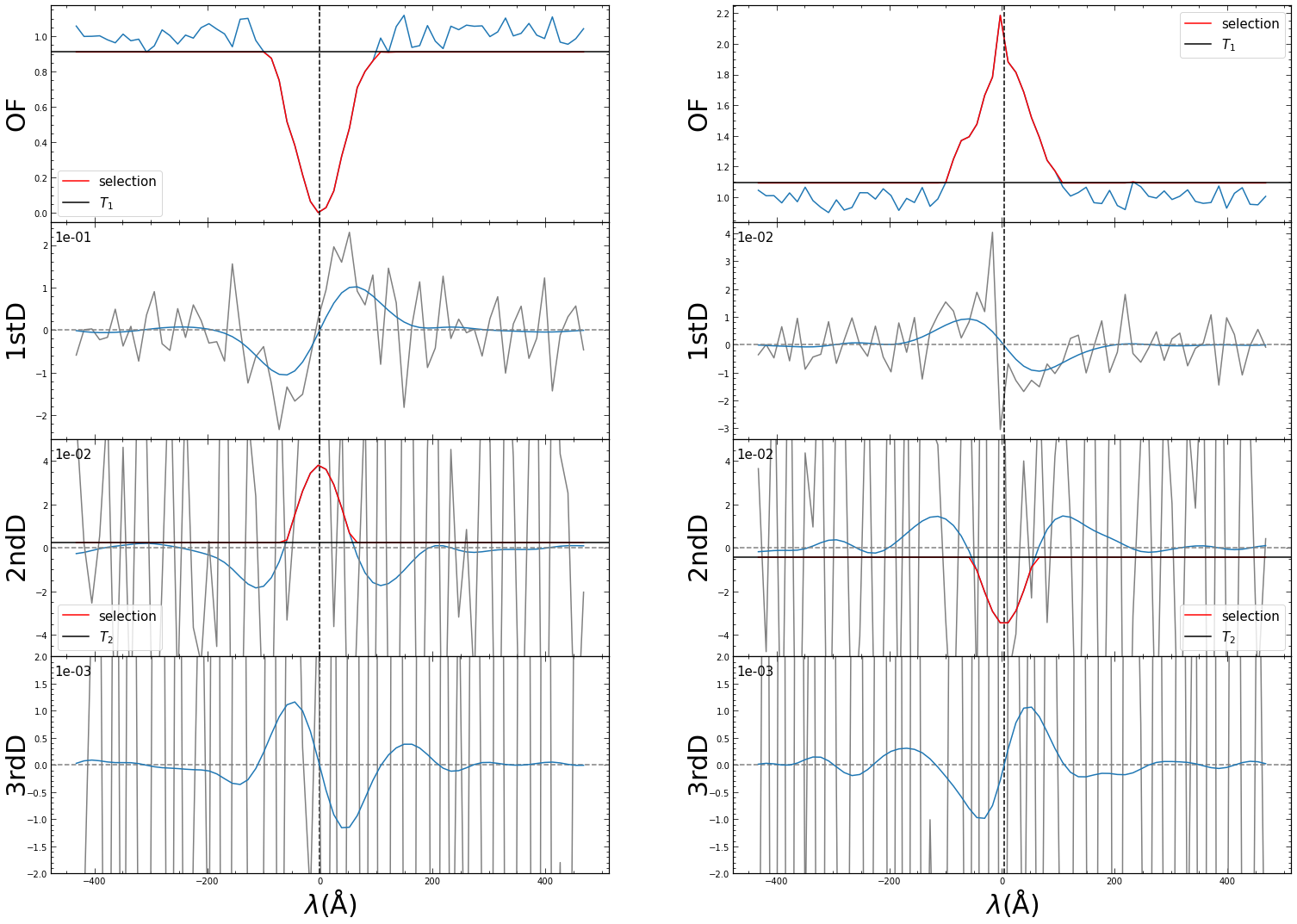}
    \caption{Original spectra and derivatives of Gaussian function with noise. The grey lines represent the derivatives obtained using a basic finite differences method, while the blue lines (in panels below the top one) represent the smoothed derivatives obtained using the $Gaussian\textunderscore filter1d$. The black horizontal lines in the top panels indicate the threshold parameter applied to the original spectrum ($T_{\rm{0}}$), and in the middle to lower panels, it indicates the threshold parameter applied to the second derivative ($T_{\rm{2}}$). The red lines highlight the selected portions.}
    \label{Fig3}
\end{figure}

We used the $Savgol\textunderscore filter$ routine from the $signal$ sub-module and $Gaussian\textunderscore filter1d$ from the $scipy$ module \citep{Jones+2001} in Python. In our testing, $Gaussian\textunderscore filter1d$ demonstrated greater stability at mid-to-low resolutions compared to $Savgol\textunderscore filter$. However, $Savgol\textunderscore filter$ exhibited superior sensitivity in high-resolution spectra. For consistency with the LAMOST-MRS spectra used in this study, we employed the $Gaussian\textunderscore filter1d$ method throughout this paper. It provided the first, second, and third derivatives of a Gaussian curve with an SNR of 20, as shown in Figure \ref{Fig3}.

A zero point in the descending part of the third derivative indicates the position of the minimum point in the second derivative, while a zero point in the ascending part of the first derivative signifies the position of the maximum point in the original (zeroth order) series. Consequently, the zero point in the descending part of the third derivative and the zero point in the ascending part of the first derivative both serve as evidence for the presence of an emission peak in the original spectrum. Conversely, a zero point in the ascending part of the third derivative and a zero point in the descending part of the first derivative indicate the existence of an absorption peak in the original spectrum.

\begin{figure}
    \centering
    \includegraphics[width=15.0cm,angle=0]{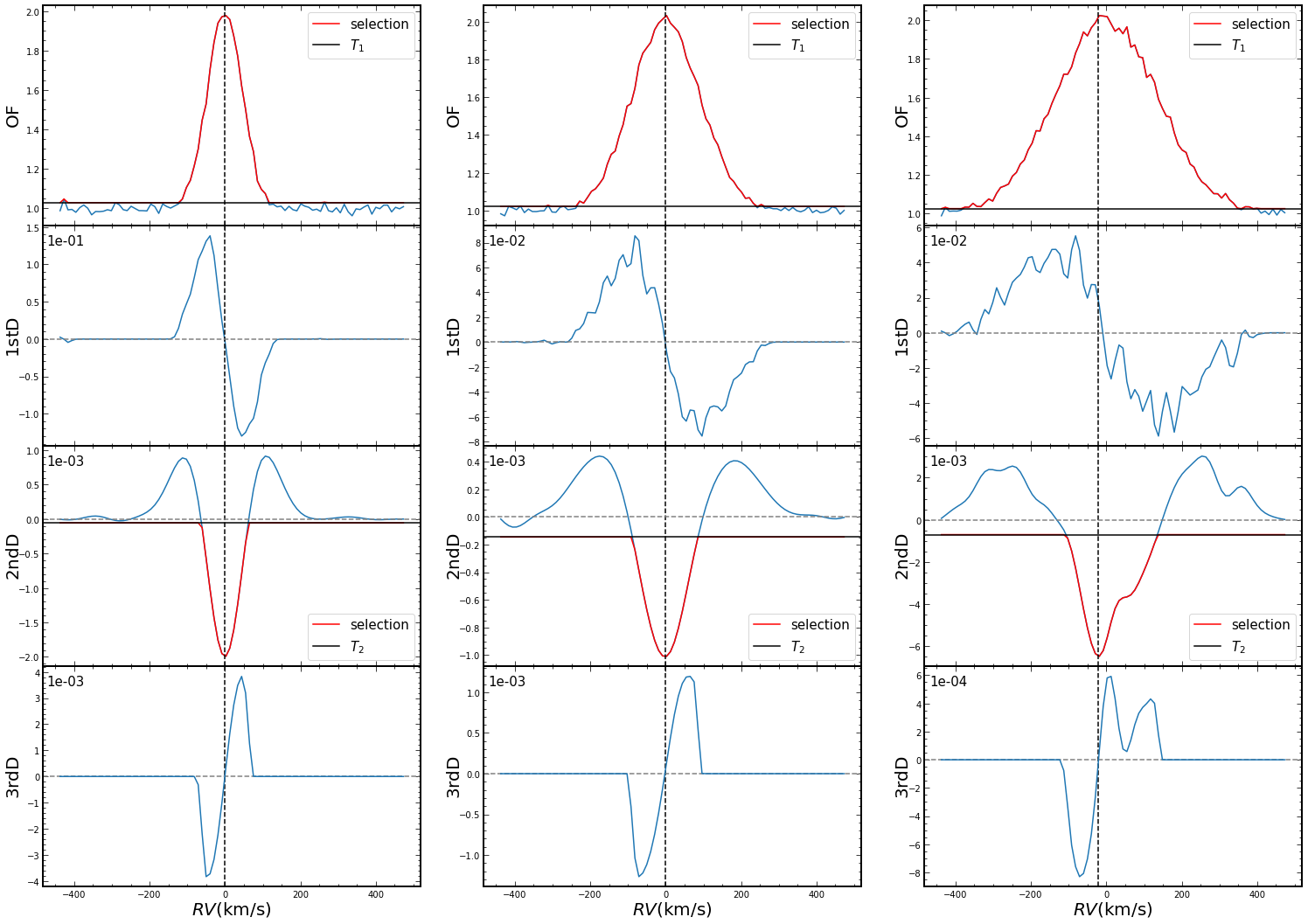}
    \caption{Emission lines created by Gaussian function with FWHM - 96.9,193.9, and 318.5 km/s.}
    \label{Fig4}
\end{figure}

\begin{figure}
    \centering
    \includegraphics[width=15.0cm,angle=0]{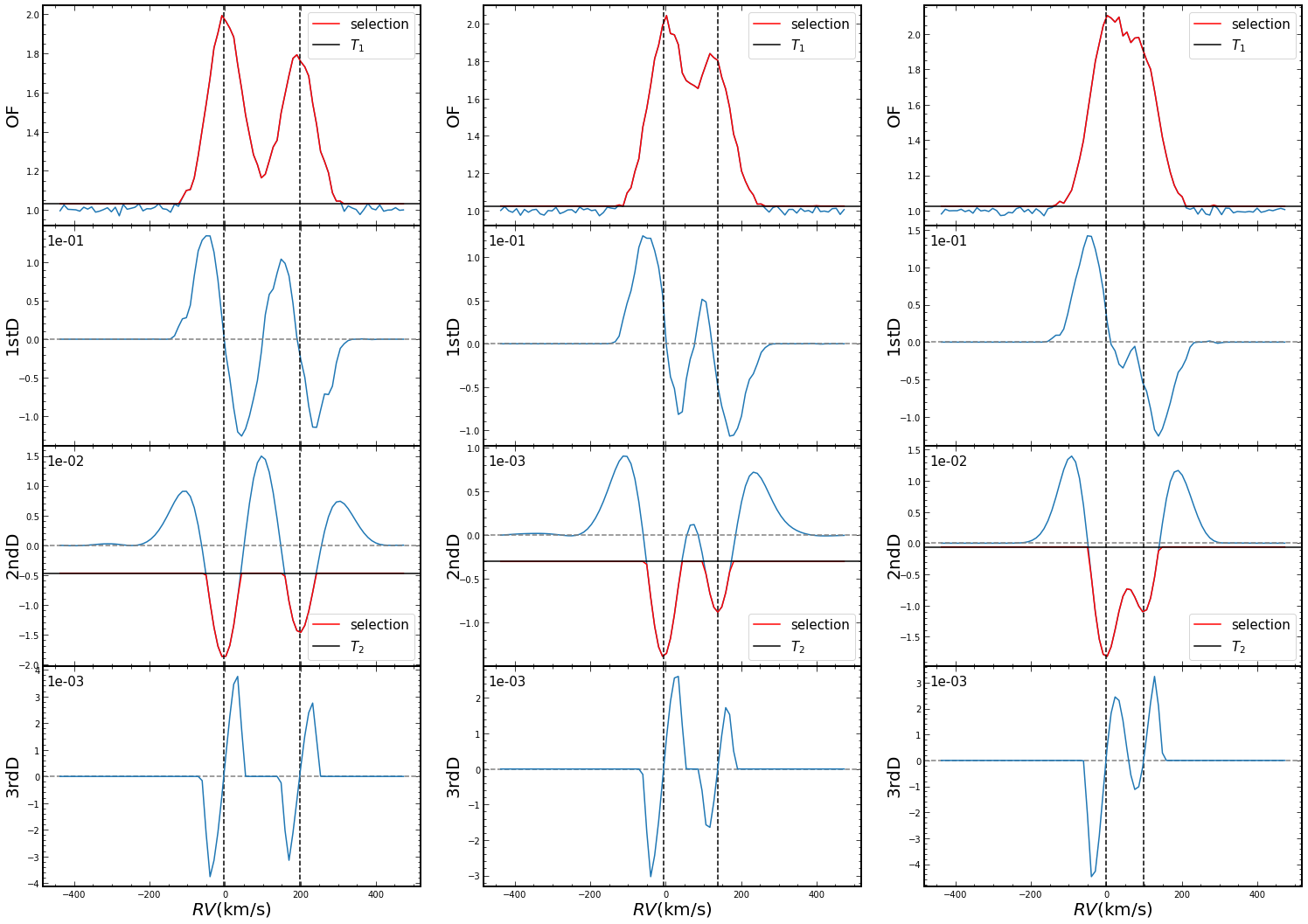}
    \caption{Double-peak spectra created by Gaussian function with internal - 193,129, and 95 km/s.}
    \label{Fig5}
\end{figure}

For emission lines, the method is applied only in the region where the original spectrum is higher than $T_{\rm{1}}$, as represented by the solid black line in the OF part of the left picture in Figure \ref{Fig3}. The selection part, represented by the solid red line, provides a declining curve in the first derivative through zero. Additionally, the selection part of the second derivative is lower than $T_{\rm{2}}$, which is also represented by the solid black line and provides a rising curve crossing through the zero point in the third derivative.

In the case of absorption lines, the selection parts are opposite to those for emission lines in both the original spectrum and the second derivative. In our tests, the absorption peak of the second derivative weakens as the Full Width Half Maximum (FWHM) of the original spectrum increases. This means it is challenging to identify the selection part in the second derivative of spectra with significantly broadened lines. Examples of our tests are shown in Figure \ref{Fig4}, and we can observe that the disadvantage of the second derivative is not evident in the original spectrum. However, the third derivative is much more sensitive than the first derivative when dealing with the resolution overlap of two signals, as demonstrated in Figure \ref{Fig5}. These characteristics of the second and third derivatives are a result of the primary benefits of derivative spectroscopy highlighted earlier in this paper - background correction and improved spectral resolution. Hence, it's essential to utilize both the first and third derivatives for detecting peaks in the original spectrum.

We employed three criteria to determine the presence of peaks in the original spectrum:
\\

1. The presence of a zero in the third derivative or first derivative.

2. The pattern of negative and positive values around the zero
point: negative values to the left of zero
\newline

\vspace{-20pt}
\hspace{5pt}
and positive values to the right of zero indicate an emission peak in the original spectrum, while the
\newline

\vspace{-20pt}
\hspace{5pt}
opposite pattern suggests an absorption peak.

3. For an emission line, the value of the original spectrum at the zero position is higher than in its
\newline

\vspace{-20pt}
\hspace{5pt}
vicinity, whereas for an absorption peak, the value is lower.
\\

Please note that Criterion 3, while ensuring the detection of emission or absorption peaks, may reduce the precision of our method. Users should apply it based on specific circumstances.

\subsection{Parameters of method}
\label{Para}

We have discussed three key parameters in our method: the width of the Gaussian kernel ($\sigma_{\rm{p}}$), the threshold for the original spectrum ($T_{\rm{1}}$), and the threshold for the second derivative ($T_{\rm{2}}$). These parameters ($\sigma_{\rm{p}}$, $T_{\rm{1}}$, and $T_{\rm{2}}$) can affect the number of detected peaks and their associated wavelengths. Once determined for a specific instrumental setup, these parameters remain constant to maintain consistent detection efficiency across the entire spectrum sample.

The $T_{\rm{1}}$ and $T_{\rm{2}}$ parameters should be set to strike a balance between avoiding noise-induced distortions when set too low and ensuring the detection of genuine but weak signal peaks when set too high. Similarly, the sigma parameter must be chosen carefully – it should not be excessively large to prevent excessive smoothing that could hinder the identification of closely spaced peaks, nor too small to minimize the impact of numerical noise resulting from successive derivatives.

To determine the optimal parameters for our method using LAMOST-MRS data, we constructed a sample set of 10,000,000 spectra with emission lines, mimicking the resolution of LAMOST-MRS. For each spectrum in this set, we varied the SNR and amplitude of the emission line, which was generated using a Gaussian function. The parameters of the Gaussian function used in the sample set are detailed in Table \ref{Tab2}. In this table, $\rm{log}(\frac{A}{\sigma_{\rm{s}}})$ represents the logarithm of the ratio between the amplitude (A) of the Gaussian function and the reciprocal of SNR ($\sigma_{\rm{s}}$), which quantifies the level of noise. $V_{\rm{0}}$ and $\sigma_{\rm{g}}$ denote the center and standard deviation of the Gaussian function, respectively.

We uniformly selected 1,000 points at $\rm{log}(\frac{A}{\sigma_{\rm{s}}})$ and generated 1,000 samples at each point, with each sample containing random noise following a normal distribution. The standard deviation of this noise was $\sigma_{\rm{s}}$, which is the reciprocal of SNR, for each of the selected points. This approach allows us to calculate the detection efficiency of our parameter settings for each amplitude, and importantly, this efficiency should remain independent of the SNR.

The $T_{\rm{1}}$ parameter, which is applied to the original spectrum and is independent of the other two parameters, was determined first. It was set to three times the standard deviation of the noise-free, smooth portion of the spectrum. This threshold eliminates 99.7$\%$ of spectra without a signal and detects signals as faint as 0.6 times the value of the reciprocal of SNR ($\sigma_{\rm{s}}$), achieving a 100$\%$ detection efficiency when the signal reaches 3 times $\sigma_{\rm{s}}$. Setting $T_{\rm{1}}$ lower would not increase detection sensitivity but would introduce many spectra without a signal into our results.

To refine our parameter selection, we drew inspiration from the methodology proposed by \cite{Merle+2017}. They assessed the precision of radial velocities computed via the Cross-Correlation Function, comparing them with the original radial velocities obtained from the Difference Of Expected (DOE) method under various $\sigma_{\rm{p}}$ parameter configurations. Our own experiments have affirmed that, within the current dataset, the $\sigma_{\rm{p}}$ parameter, spanning from 0.5 Å to 2.5 Å, proves to be robust for our methodology.

The parameters $\sigma_{\rm{p}}$ and $T_{\rm{1}}$ are interconnected and collectively impact the selection criteria for the second derivative. To determine the $\sigma_{\rm{p}}$ parameter, we employed a marginalization approach, ultimately setting it at 0.61 Å. Subsequently, we determined the $T_{\rm{2}}$ parameter to be 3.5 times the standard deviation. With these settings, our method can detect signals with an amplitude as low as 1 times the standard deviation in the second derivative for signals with a half-width of 2.1 Å. Similarly, for signals with a half-width of 6.3 Å, these parameters can detect signals with an amplitude as low as 3 times the standard deviation while eliminating spectra without signals in nearly 99.85$\%$ of cases. Lower $T_{\rm{2}}$ settings would fail to detect lower-amplitude signals while significantly increasing the number of detected spectra without signals.

In our testing, all three parameters for LAMOST-MRS spectra within the specified range in Table \ref{Tab1} have proven to be reliable. However, it's important to note that the $\sigma_{\rm{p}}$ and $T_{\rm{2}}$ parameters are interrelated. Adjusting one may require fine-tuning the other.




\begin{table}
\centering
\begin{minipage}[t]{0.35\linewidth} 
\centering

\begin{tabularx}{\linewidth}{XX}
    \hline\hline
        SNR &  [10,100](dex=10) \\
        $\rm{log}(\frac{A}{\sigma_{\rm{s}}})$ & [-1,1] \\
        $V_{\rm{0}}$ & 0 \\
        $\sigma_{\rm{g}}$ & 1 \\
    \hline
    \end{tabularx}
\caption{Parameter space of the dataset.}
\label{Tab2}

\end{minipage}
\hfill 
\begin{minipage}[t]{0.55\linewidth} 
\centering
\begin{tabularx}{\linewidth}{XXX}
    \hline\hline
    Para & LAMOST-MRS & LAMOST-LRS  \\ \hline
    $\sigma_{\rm{p}}$ & 0.61 - 0.76 Å & 0.6 Å\\
    $T_{\rm{1}}$ & 3-3.3 &  3  \\
    $T_{\rm{2}}$ & 3.5-4.5 & 3-4.4 \\\hline
\end{tabularx}
\caption{Setups used in LAMOST and the associated estimated parameters}
\label{Tab1}

\end{minipage}
\end{table}

\begin{figure}
    \centering
    \includegraphics[width=15.0cm,angle=0]{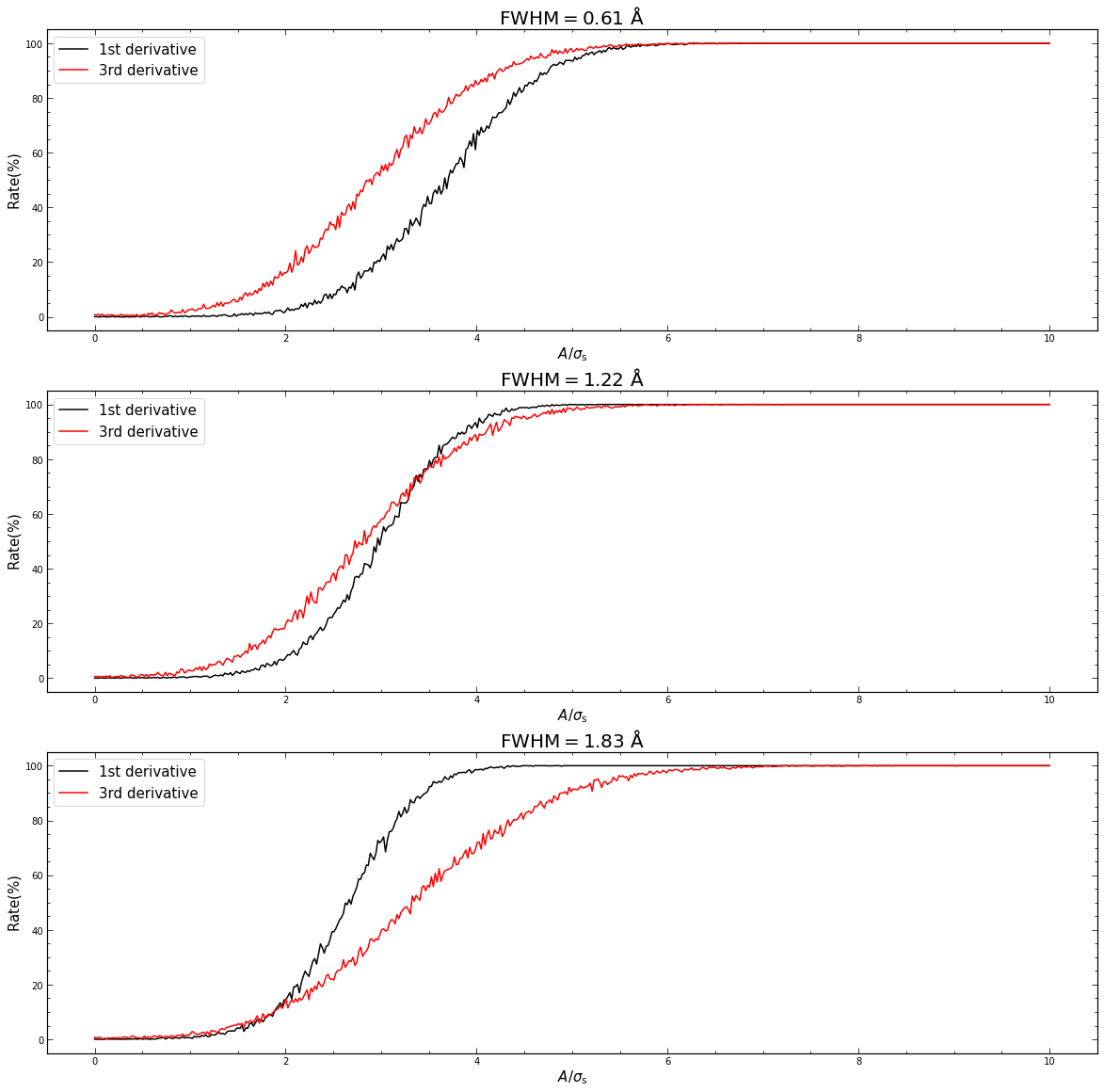}
    \caption{Detect efficiency of the first derivative and third derivative at FWHM 0.61,1.22 and 1.83 Å.}
    \label{Fig6}
\end{figure}

Under the current parameter settings, we have observed that when the full width at half maximum (FWHM) is 1.22 Å, the detection efficiency of the first derivative and third derivative are nearly identical. The subtle difference between them lies in the fact that the third derivative is better at detecting faint signals below 2.5 times the standard deviation ($\sigma_{\rm{s}}$), while the first derivative excels at detecting faint signals above 2.5 times $\sigma_{\rm{s}}$. When the FWHM is below 1.22 Å, the first derivative's detection efficiency is lower than that of the third derivative. Conversely, when the FWHM is above 1.22 Å, the first derivative's detection efficiency surpasses that of the third derivative. You can see the detection efficiencies of the first and third derivatives at FWHM values of 0.61, 1.22, and 1.83 Å in Figure \ref{Fig6}. The detailed reasons for these differing detection efficiencies are explained in Section \ref{Med}.


\subsection{Test at synthesis spectra}
\label{Tes}

The sample set of artificial spectra was introduced in the previous section. The detection efficiency of the derivative spectroscopy (DS) method for each amplitude is depicted as a black solid line in Figure \ref{Fig8}. Across the four panels with SNR values ranging from 20(a) to 80(d), the detection efficiency consistently starts to rise at 0.5 times the value of $\sigma_{\rm{s}}$, which is 15.5$\%$ and reaches 97.9$\%$ at 3.5 times $\sigma_{\rm{s}}$. We also replicated the Gaussian fitting detection method and conducted comparative tests using the same dataset.

\begin{figure}
    \centering
    \includegraphics[width=15.0cm, angle=0]{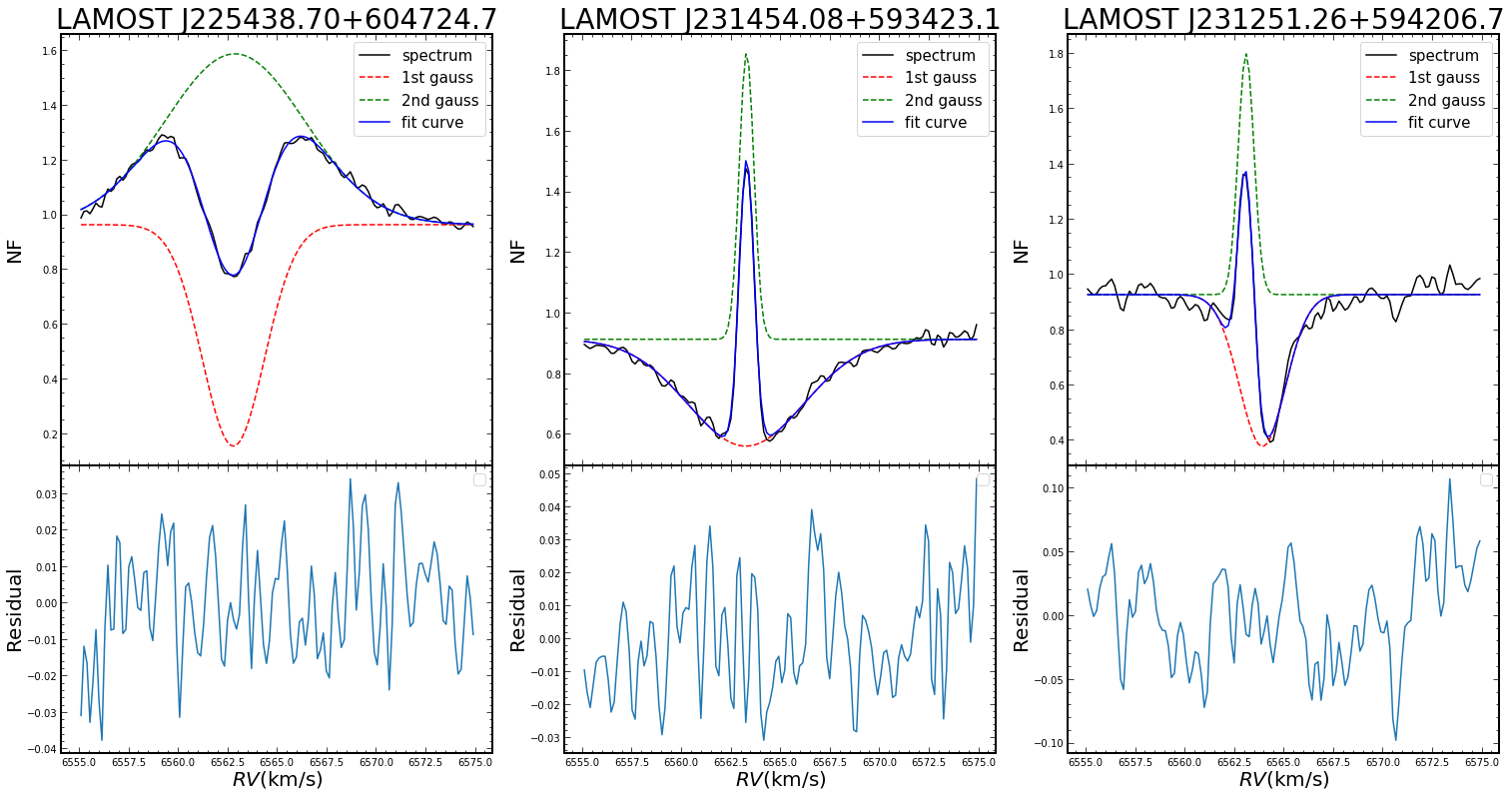}
    \caption{Spectra from the LAMOST-MRS dataset were selected as examples, and Gaussian functions were applied to fit the lines of these spectra in terms of normalized flux(NF).}
    \label{Fig7}
\end{figure}

\begin{figure}
    \centering
    \includegraphics[width=15.0cm, angle=0]{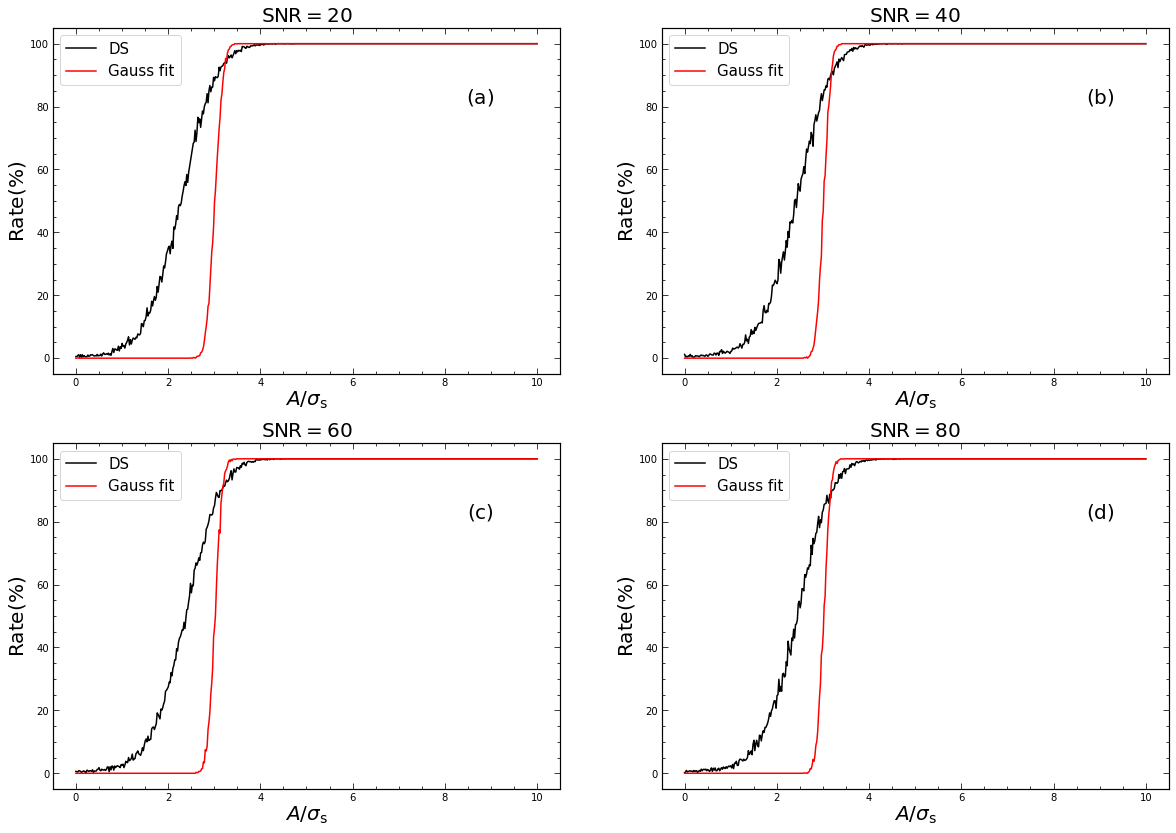}
    \caption{Detect efficiencies of DS and Gaussian fit.}
    \label{Fig8}
\end{figure}

Gaussian Fit is a robust method for detecting the different components of a complicated spectrum structure by fitting the profile of a spectral line with Gaussian functions. In Figure \ref{Fig7}, we used two Gaussian functions to fit three $\rm{H\alpha}$ lines from LAMOST-MRS spectra. Each line is divided into two components: the green dotted line represents the emission line, and the red dotted line represents the absorption line. The blue solid line is the fitted result, while the black solid line represents the original line. The fitting residuals for each spectrum are all lower than 0.1.

The detection efficiency of the Gaussian fit for each amplitude is also shown in Figure \ref{Fig8}. Similar to DS, it reaches 99.9$\%$ at 3.5 times the $\sigma_{\rm{s}}$ but has difficulty detecting signals lower than 2.7 times, which is 17$\%$.

\section{Test at LAMOST-MRS}
\label{sec:tes}

\subsection{Data selection}
\label{Dat}

The LAMOST telescope, also known as the Guo Shou Jing Telescope, is a special reflecting Schmidt telescope\citep{Cui+2012,Zhao+2012}. The Medium-Resolution Spectroscopic Survey of LAMOST (LAMOST-MRS) offers spectra in the wavelength ranges covered by the blue and red arms, which are 4950-5350 Å and 6300-6800 Å, respectively\citep{Hou+2018}. To detect emission lines and their associated wavelengths, we have specifically chosen the $\rm{H\alpha}$ profile located in the red arm, as it exhibits the most significant profile characteristics.

It's important to note that spectra with low SNR can make it challenging to reliably detect peaks. Consequently, we have selected only the \rm{$H\alpha$} lines in the red arm with an SNR greater than 10 to form our sample set. To ensure consistency and facilitate further analysis, we have applied a normalization procedure to all samples using the sub-module $normalization\textunderscore spectrum\textunderscore spline$ of the $laspec$ module. This normalization process is a standard procedure for spectra from LAMOST-MRS data \citep[e.g.][]{Zhang+2021}. The normalization process involves the following steps:
\\

1. Each spectrum is divided evenly into 10 bins based on wavelength.

2. Quadratic spline interpolation is applied to these 10 bins using the median flux values.
\\

This procedure helps to standardize and prepare the spectra for further analysis, making them more amenable to peak detection and other analytical techniques.

\begin{figure}
    \centering
    \includegraphics[width=15.0cm, angle=0]{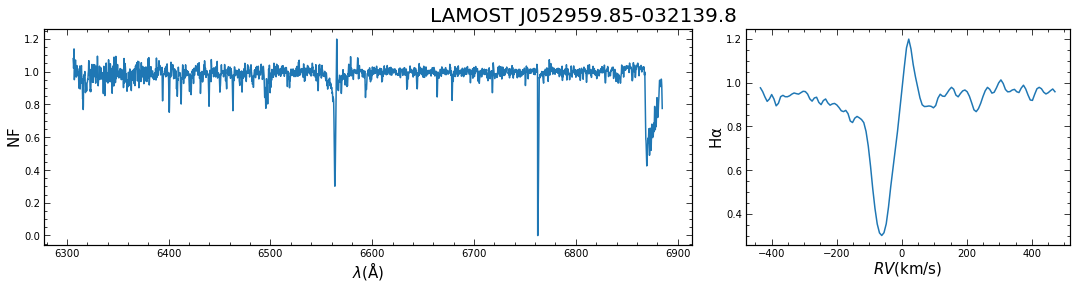}
    \caption{Examples of the normalized spectrum of LAMOST-MRS and its profile of $\rm{H\alpha}$ line.}
    \label{Fig9}
\end{figure}

Fig \ref{Fig9} displays a normalized red-arm spectrum from LAMOST-MRS on the left, and on the right, it shows the radial velocity profile of the $\rm{H\alpha}$ line. The wavelength range in the right panel is the specific region of interest.

\subsection{Results}
\label{Res}
We conducted our method's testing on a randomly selected subset of 579,680 coadded spectra from LAMOST. These spectra originated from 249,324 stars cross-matched with Gaia data. Derivative spectroscopy identified 23,804 spectra with either emission lines or double absorption peaks. We excluded 7,031 spectra that exhibited very strong negative values or zero flux values around $\rm{H\alpha}$ in the spectra. Additionally, we excluded 144 spectra, which accounted for less than 0.86$\%$ of the total, that initially appeared to have absorption profiles but were misidentified as emissions during the first step of data reduction, likely due to noise or cosmic rays.

In presenting our results, we often use a velocity scale for the wavelength values, with zero centered at the $\rm{H\alpha}$ line.

\subsubsection{Morphological classification}
\label{Mor}
The DS method described in the preceding section is primarily a morphological analysis of spectra. It doesn't directly pertain to the underlying physics of the observed object. Instead, it extracts information about the wavelength and amplitude of emission and absorption lines from the derivative analysis.

As a result, we can use a limited set of parameters to classify all 10,963 analyzed objects, along with their 16,629 spectra, into meaningful morphological categories, and potentially even into physical categories.

In our classification, we divided the 16,629 spectra into three main classes: 1. Emission peak above the pseudo-continuum line - 34.6$\%$ 2.Emission peak below pseudo-continuum line - 64.9$\%$ 3.Emission peak between pseudo-continuum line - 1.5$\%$:
\\\\
1. Emission Peak Above Pseudo-continuum Line (Class I): This class comprises four sub-types (Type 1.1, Type 1.2, Type 1.3, Type 1.4, Type 1.5). The presence of an emission peak above the continuum strongly suggests the possibility of the star being an emission-line star.
\\
\newline

\vspace{-20pt}
\hspace{5pt}
(a) Single-peak (Type 1.1):
- Profiles of this type are characterized by a prominent emission peak that
\newline

\vspace{-20pt}
\hspace{5pt}
stands out above the continuum. This distinct feature makes them easily identifiable in our analysis.
\newline

\vspace{-20pt}
\hspace{5pt}
(b) Double-peak (Type 1.2):
- These profiles also display explicit emission characteristics. The key
\newline

\vspace{-20pt}
\hspace{5pt}
difference from Type 1.1 is that Type 1.2 has two emission peaks around the $\rm{H\alpha}$ line. Most of them
\newline

\vspace{-20pt}
\hspace{5pt}
in our sample exhibit wider winds compared to other spectra in Class I.
\newline

\vspace{-20pt}
\hspace{5pt}
(c) Emission-peak within an Absorption Line (Type 1.3):
- Unlike the previous two types, this profile
\newline

\vspace{-20pt}
\hspace{5pt}
type features a wide, shallow absorption component overlaying the emission line.
\newline

\vspace{-20pt}
\hspace{5pt}
(d) P Cygni Profile (Type 1.4):
- This profile type exhibits an absorption component to the left of the
\newline

\vspace{-20pt}
\hspace{5pt}
$\rm{H\alpha}$ emission line, indicating an expanding envelope surrounding the central star.
\newline

\vspace{-20pt}
\hspace{5pt}
(e) Inverse P Cygni Profile (Type 1.5):
- Unlike Type 1.3, the absorption component in this type shifts
\newline

\vspace{-20pt}
\hspace{5pt}

\begin{figure}[H]
    \centering

    \begin{subfigure}{0.4\textwidth}
        \centering
        \includegraphics[width=\linewidth]{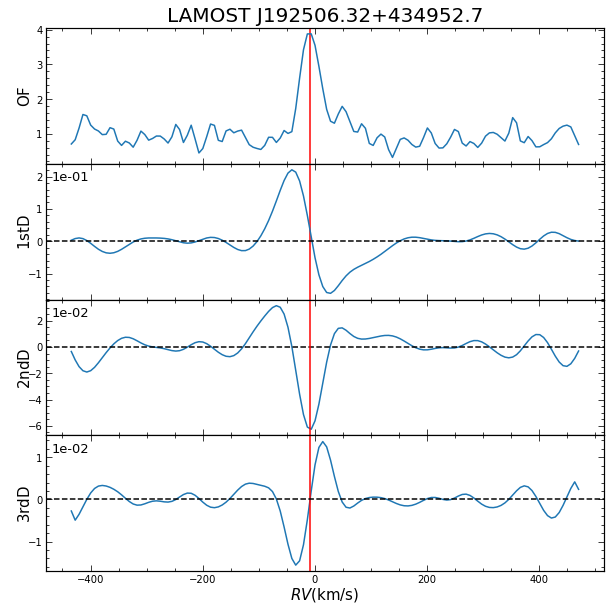}
        \caption{Type1.1}
        \label{fig:sub1}
    \end{subfigure}
    \hfill
    \begin{subfigure}{0.4\textwidth}
        \centering
        \includegraphics[width=\linewidth]{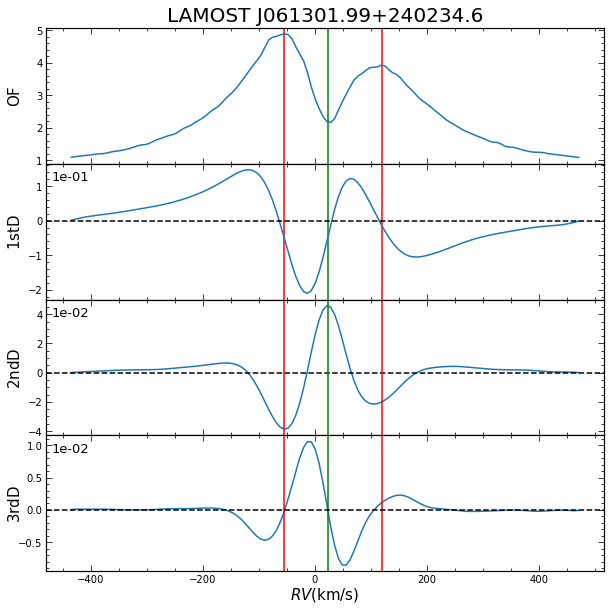}
        \caption{Type1.2}
        \label{fig:sub2}
    \end{subfigure}
    \hfill

    \begin{subfigure}{0.4\textwidth}
        \centering
        \includegraphics[width=\linewidth]{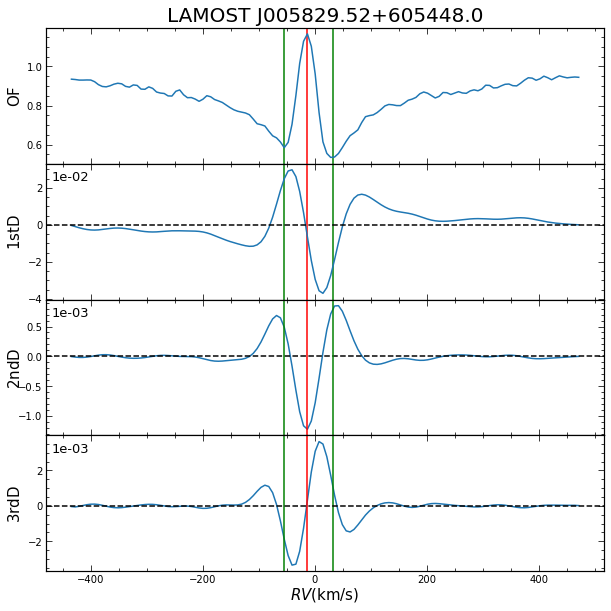}
        \caption{Type1.3}
        \label{fig:sub3}
    \end{subfigure}
    \hfill

    \begin{subfigure}{0.4\textwidth}
        \centering
        \includegraphics[width=\linewidth]{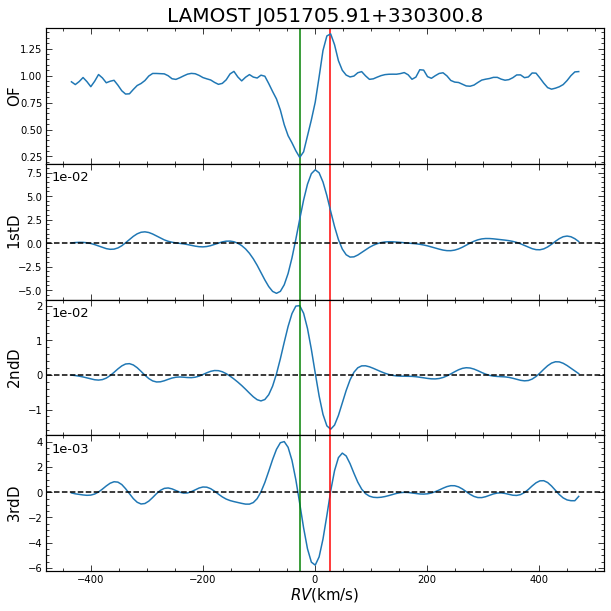}
        \caption{Type1.4}
        \label{fig:sub4}
    \end{subfigure}
    \hfill
    \begin{subfigure}{0.4\textwidth}
        \centering
        \includegraphics[width=\linewidth]{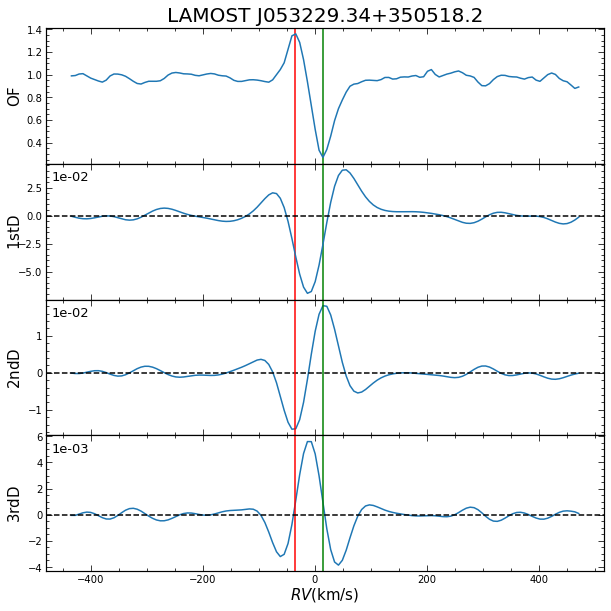}
        \caption{Type1.5}
        \label{fig:sub5}
    \end{subfigure}

    \caption{$\rm{H\alpha}$ emission profiles of ClassI and its first, second, and third derivative. The red and green solid lines respectively mark the emission peaks and absorption peaks of the original spectrum. Please note that the peak markers are derived from the selection of first and third derivatives by $T_{\rm 1}, T_{\rm 2}$, and they do not align perfectly with the zero-crossings of the unfiltered derivative spectrum.}
    \label{Fig10}
\end{figure}

to the right of the $\rm{H\alpha}$ emission line. This red-shifted absorption component indicates a contracting
\newline

\vspace{-20pt}
\hspace{5pt}
envelope surrounding the star.
\\\\

2. Emission-peak below the pseudo-continuum line(ClassII): This class encompasses three sub-types (Type 2.1, Type 2.2, and Type 2.3). This classification suggests that the star may exhibit an emission-line profile with either deep absorption or multiple absorption peaks, making it challenging to discern through morphological analysis.
\\
\newline

\vspace{-20pt}
\hspace{5pt}
(a) Type 2.1: Sharp emission-peak profiles in this category feature an emission peak identified by
\newline

\vspace{-20pt}
\hspace{5pt}
derivative spectroscopy, situated at the center of a deep absorption component. The peak of
\newline

\vspace{-20pt}
\hspace{5pt}
the emission line remains below the pseudo-continuum line.
\newline

\vspace{-20pt}
\hspace{5pt}
(b) Type 2.2: This type is similar to Type 2.1, with the distinction that the emission peak is scarcely
\newline

\vspace{-20pt}
\hspace{5pt}
discernible by the third derivative. Instead, two clearly defined absorption peaks are evident around
\newline

\vspace{-20pt}
\hspace{5pt}
the $\rm{H\alpha}$ line.
\newline

\vspace{-20pt}
\hspace{5pt}
(c) Type 2.3: Profiles in this category also possess an emission peak detectable by derivative
\newline

\vspace{-20pt}
\hspace{5pt}
spectroscopy. However, in Type 2.3, the emission peaks are shifted to one side of the absorption,
\newline

\vspace{-20pt}
\hspace{5pt}
resulting in the location of an absorption peak on only one side of the emission peak.
\\\\
3. Emission peak between continuum(ClassIII): Unlike the other two classes, this class exhibits an emission peak positioned between the continuum levels. Here, the emission peak is higher on the red side of the $\rm{H\alpha}$ line and lower on the blue side, resulting in a noticeable jump around the $\rm{H\alpha}$ line.
\\

\begin{figure}
    \centering

    \begin{subfigure}{0.4\textwidth}
        \centering
        \includegraphics[width=\linewidth]{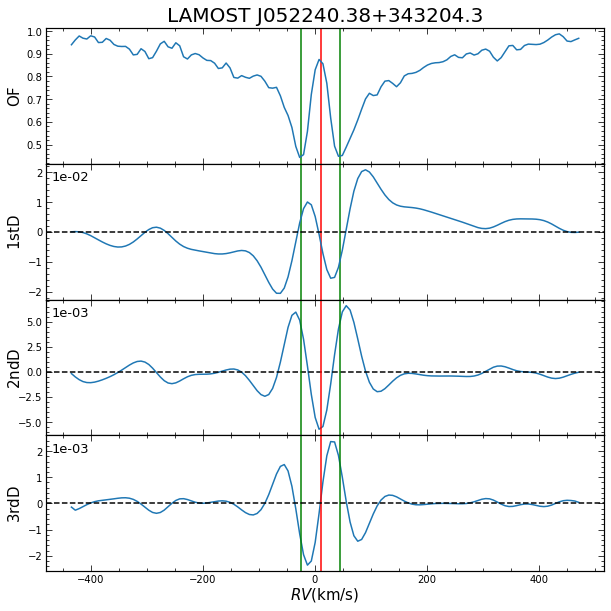}
        \caption{Type2.1}
        \label{fig:11a}
    \end{subfigure}
    \hfill
    \begin{subfigure}{0.4\textwidth}
        \centering
        \includegraphics[width=\linewidth]{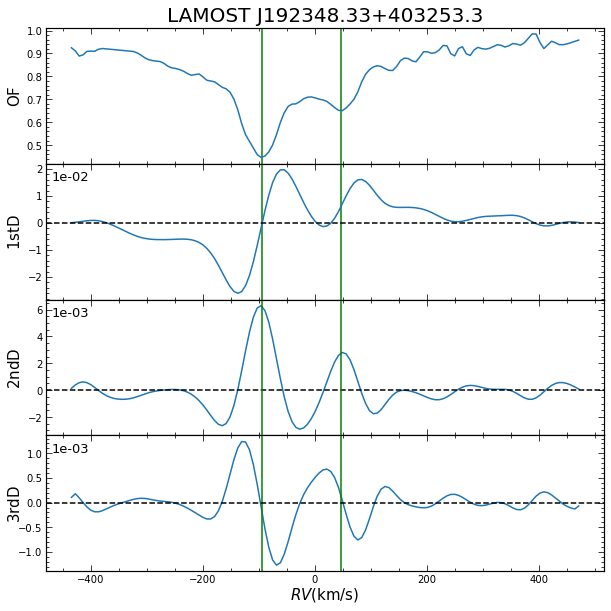}
        \caption{Type2.2}
        \label{fig:11b}
    \end{subfigure}
    \hfill

    \begin{subfigure}{0.4\textwidth}
        \centering
        \includegraphics[width=\linewidth]{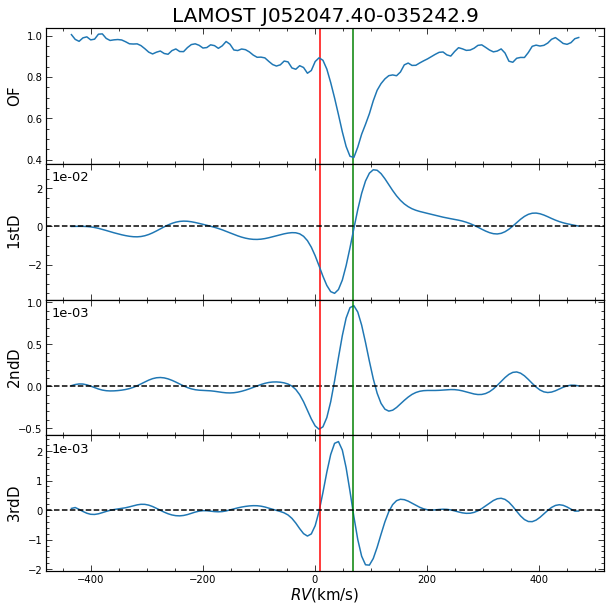}
        \caption{Type2.3}
        \label{fig:11c}
    \end{subfigure}

    \caption{Same as Fig\ref{Fig10} but for ClassII.}
    \label{Fig11}
\end{figure}

\begin{figure}
    \centering

    \begin{subfigure}{0.7\textwidth}
        \centering
        \includegraphics[width=\linewidth]{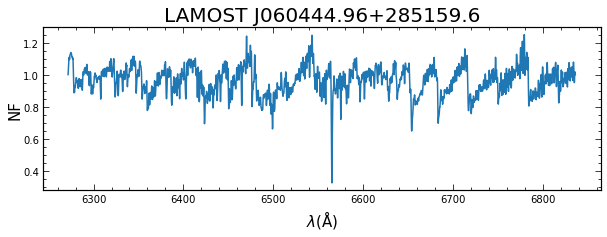}
        \caption{}
        \label{fig:12a}
    \end{subfigure}
    \hfill
    \begin{subfigure}{0.7\textwidth}
        \centering
        \includegraphics[width=\linewidth]{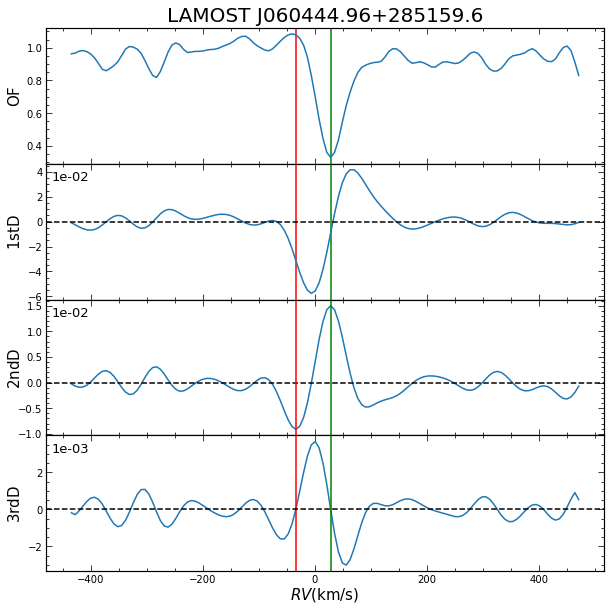}
        \caption{}
        \label{fig:12b}
    \end{subfigure}

    \caption{Example(b) of Class III as Fig \ref{Fig10} and its normalized spectrum(a) of red arm.}
    \label{Fig12}

\end{figure}

For each of the nine subclasses, we plotted the normalized flux spectra in the $\rm{H\alpha}$ band, along with their first, second, and third derivative spectra at Fig \ref{Fig10},\ref{Fig11}, and \ref{Fig12}. The spectral classification is based on the morphological features of the $\rm{H\alpha}$ line profiles, as mentioned earlier. It is important to note that this classification may not necessarily be related to the underlying physics of the stars. Nevertheless, we still hope to impose certain constraints on the corresponding stars based on the classification of the spectral line profiles.

For example, type 1.2 may stem from phenomena like stellar jets and accretion disks, whereas type 1.4 and 1.5 are indicative of the stellar envelope undergoing outward and inward motion, respectively. Class II may arise from emission lines within absorption lines or SB2 binary stars. The jumps in class III spectra may be attributed to molecular absorption bands in cooler stars. We provide the red-arm normalized flux spectra for this type in Fig \ref{fig:12a}, which offers a more intuitive view.

\subsubsection{Classification Criteria}

We established nine sub-classes along with the conditions for their assignment, detailed in Table \ref{Tab3}. This classification system utilizes eight parameters $(N_{\rm{E}},N_{\rm{A}},NF_{\rm{E}},RV_{\rm{E}},NF_{\rm{A}},RV_{\rm{A}},NF_{\rm{r}},NF_{\rm{b}})$, which are explained below. Their values are directly derived from our method:
\\

1.$N_{\rm{E}}$ and $N_{\rm{A}}$ represent the number of emission and absorption peaks around $\rm{H\alpha}$ in the spectrum. These
\newline

\vspace{-20pt}
\hspace{5pt}
peaks are identified by locating the zero-crossings in the rising and declining parts of the first and
\newline

\vspace{-20pt}
\hspace{5pt}
third derivatives of the original spectrum.

2.$NF_{\rm{E}}$ and $RV_{\rm{E}}$ stand for the amplitude and radial velocity of emission peaks respectively. In cases
\newline

\vspace{-20pt}
\hspace{5pt}
where $N_{\rm{E}}$ is greater than 1, these parameters are denoted as $NF_{\rm{E1}},NF_{\rm{E2}},NF_{\rm{E3}}\cdots$, as well as
\newline

\vspace{-20pt}
\hspace{5pt}
$RV_{\rm{E1}},RV_{\rm{E2}},RV_{\rm{E3}}\cdots$, following the sequence from red to blue. The amplitude and radial velocity of
\newline

\vspace{-20pt}
\hspace{5pt}
emission peaks are associated with their normalized flux and wavelength. The normalized flux can be
\newline

\vspace{-20pt}
\hspace{5pt}
easily obtained by locating the zero-crossings in the rising part of the third derivative spectrum and
\newline

\vspace{-20pt}
\hspace{5pt}
applying them to the original spectrum.

3.$NF_{\rm{A}}$ and $RV_{\rm{A}}$ represent the amplitude and radial velocity of absorption peaks. Similarly, these
\newline

\vspace{-20pt}
\hspace{5pt}
parameters are labeled as $NF_{\rm{A1}},NF_{\rm{A2}},NF_{\rm{A3}}\cdots$ and $RV_{\rm{A1}},RV_{\rm{A2}},RV_{\rm{A3}}\cdots$, derived from the
\newline

\vspace{-20pt}
\hspace{5pt}
zero-crossings in the declining part of the third derivative spectrum.

4.$NF_{\rm{r}}$ and $NF_{\rm{b}}$ are the median values of the red and blue bands of the $\rm{H\alpha}$ line, each with three times
\newline

\vspace{-20pt}
\hspace{5pt}
the standard deviation.
\\

\begin{table}[]
    \centering
    \begin{tabular}{cccccc}
    \hline
    \hline
    Class &Type & Condition I & Condition II & Condition III & Condition IV\\ \hline
    Class I & 1.1 & $NF_{\rm{E}} > NF_{\rm{r,b}}$  & $ N_{\rm{E}} = 1 $ & $N_{\rm{A}} = 0$\\
    &1.2 & $NF_{\rm{E}} > NF_{\rm{r,b}} $ & $N_{\rm{E}} = 2$ & $N_{\rm{A}} = 0$\\
    &1.3 & $ NF_{\rm{E}} > NF_{\rm{r,b}} $ & $N_{\rm{E}} = 1$ & $N_{\rm{A}} = 2$ & $RV_{\rm{A1}} < RV_{\rm{E}} < RV_{\rm{A2}}$\\
    &1.4 & $NF_{\rm{E}} > NF_{\rm{r,b}} $ & $N_{\rm{E}} = 1$ & $N_{\rm{A}} = 1$ &  $ \rm{183}\ \rm{km/s} > RV_{\rm{E}} - RV_{\rm{A}} > 0$ \\
    &1.5 & $NF_{\rm{E}} > NF_{\rm{r,b}} $ & $N_{\rm{E}} = 1$ & $N_{\rm{A}} = 1$ & $0 > RV_{\rm{E}} - RV_{\rm{A}} > \rm{-183}\ \rm{km/s}$\\
    Class II& 2.1 & $NF_{\rm{A,E}} < NF_{\rm{r,b}}$ & $N_{\rm{E}} = 1$ & $N_{\rm{A}} = 2$ & $RV_{\rm{A1}} < RV_{\rm{E}} < RV_{\rm{A2}}$ \\
    &2.2 & $NF_{\rm{A,E}} < NF_{\rm{r,b}} $ & $N_{\rm{E}} = 0$ & $N_{\rm{A}} = 2$ & \\
    &2.3 & $NF_{\rm{A,E}} < NF_{\rm{r,b}} $ & $N_{\rm{E}} = 1$ & $N_{\rm{A}} = 1$ & $\rm{183}\ \rm{km/s} > |RV_{\rm{A}} - RV_{\rm{E}}| > 0$\\
    Class III & 3 & $NF_{\rm{b}} > NF_{\rm{E}} > NF_{\rm{r}}$ & $N_{\rm{E}} = 1$ & $ N_{\rm{A}} = 1 $ & $RV_{\rm{A}} - RV_{\rm{E}} > 0$ \\ \hline
    \end{tabular}
    \caption{Scheme of morphological classification of 9 sub-classes.}
    \label{Tab3}
\end{table}

\section{Discussion}
\label{sect:Dis}

DS is one of the possibilities for morphological analysis of weak signals of profiles of spectral lines, which provides a technology to enhance the resolution of the spectrum and divide the weak signal from useless background. The method we introduced here was used to first detect profiles of $\rm{H\alpha}$ line and then classify them based on a few parameters of DS, although this method could be used for other spectral lines in a similar way. We compared DS with Gaussian fit at artificial spectra, which indicates DS is more sensitive at a weak signal and enables us to detect the signal with an amplitude that is lower than 3 times of $\sigma_{\rm{s}}$.

\begin{figure}
    \centering

    \begin{subfigure}{0.8\textwidth}
        \centering
        \includegraphics[width=\linewidth]{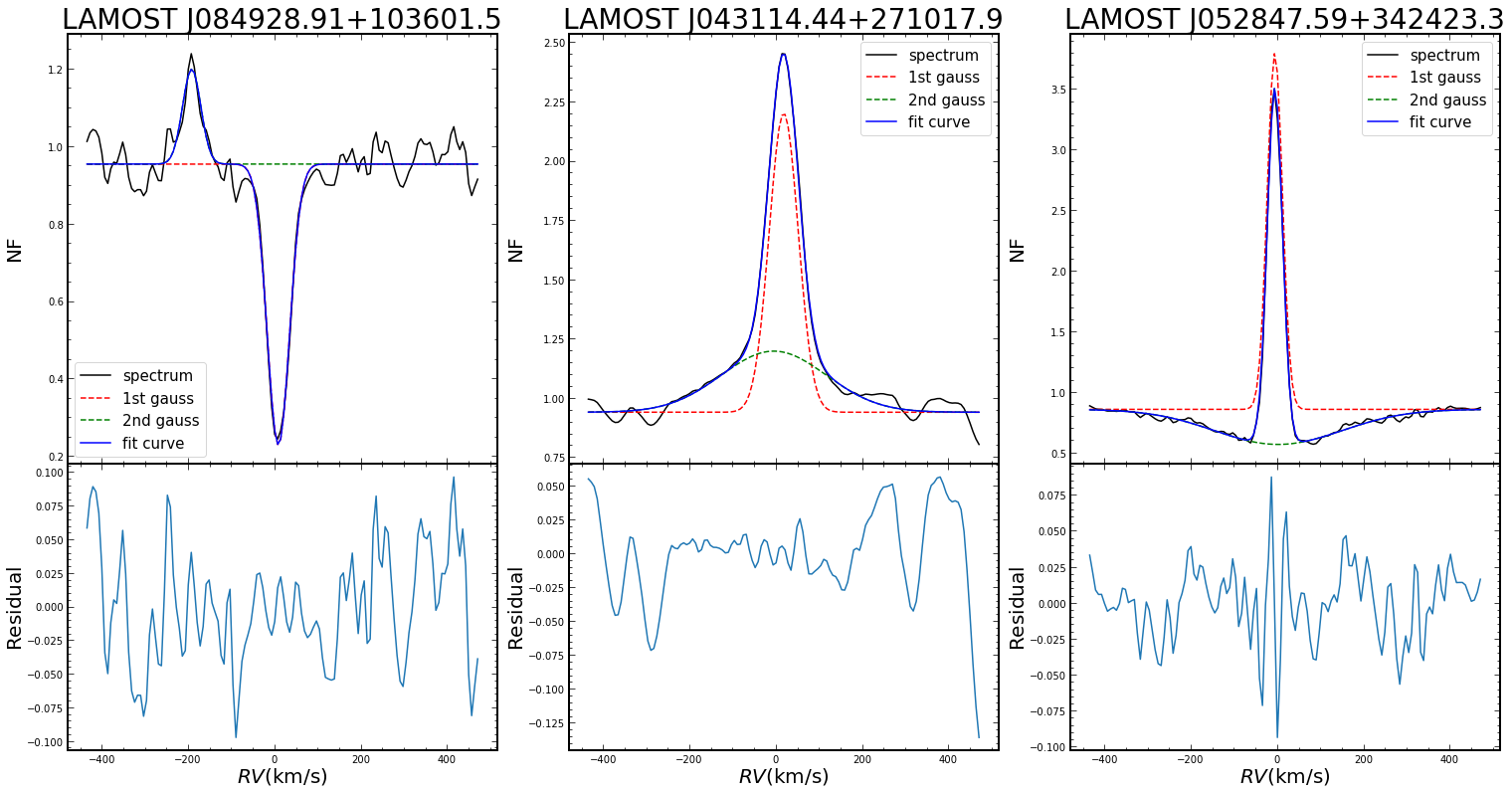}
        \caption{Gaussian fit}
        \label{fig:13a}
    \end{subfigure}
    \hfill
    \begin{subfigure}{0.8\textwidth}
        \centering
        \includegraphics[width=\linewidth]{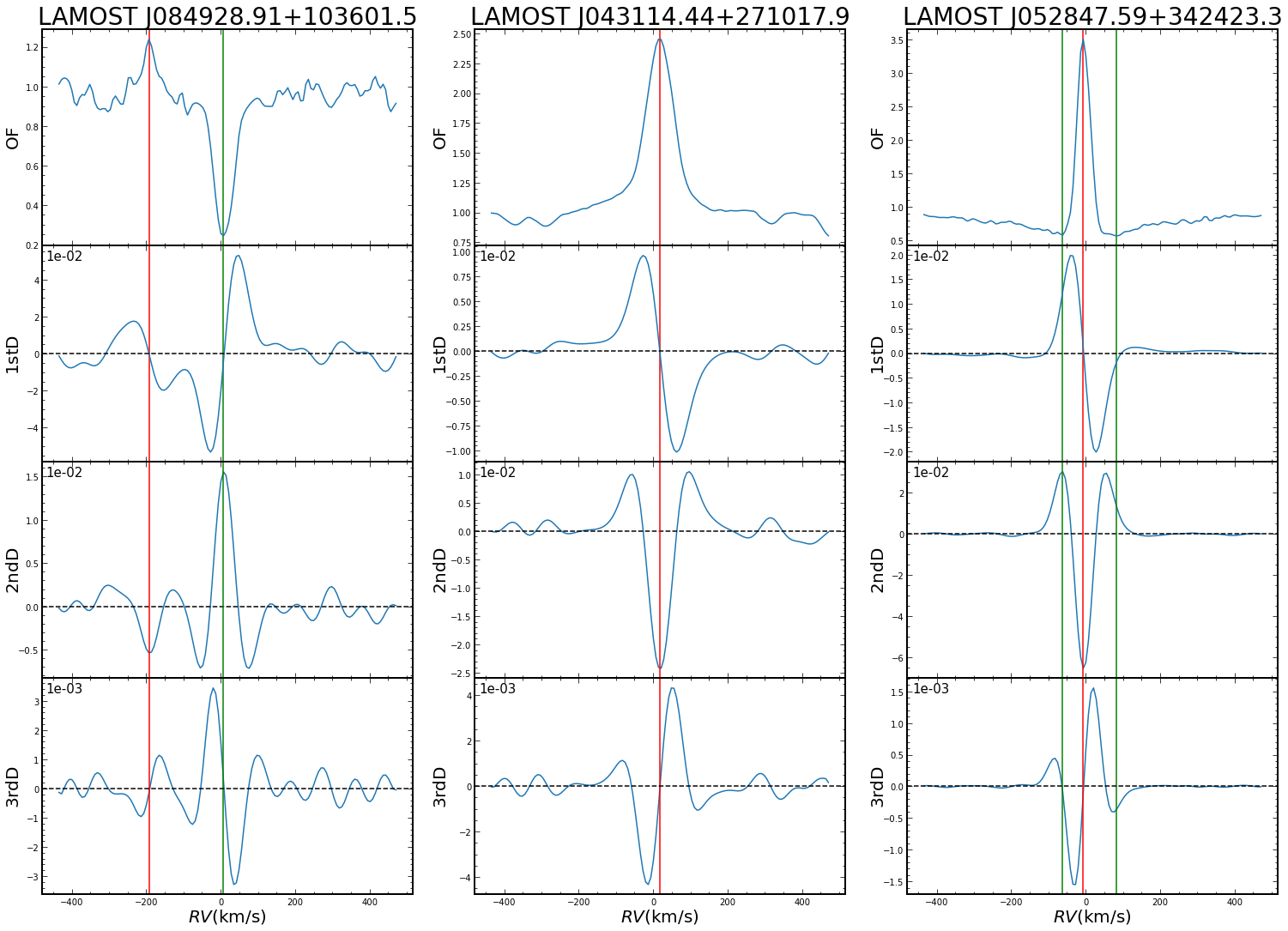}
        \caption{DS}
        \label{fig:13b}
    \end{subfigure}

    \caption{Gaussian fit and DS of J084928.91+103601.5,J043114.44+271017.9,J052847.59+342423.3}
    \label{Fig13}

\end{figure}

\begin{figure}
    \centering

    \begin{subfigure}{0.8\textwidth}
        \centering
        \includegraphics[width=\linewidth]{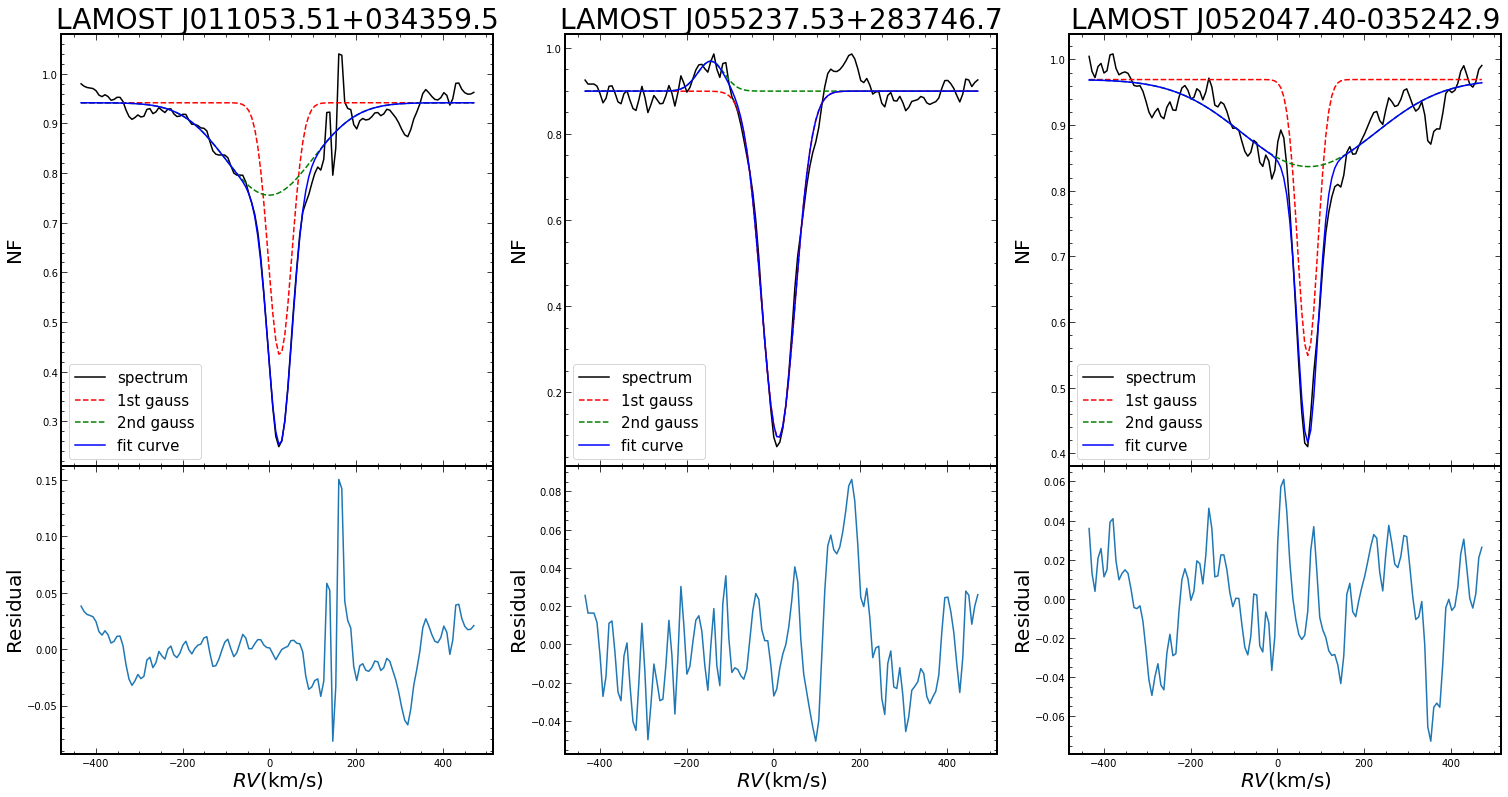}
        \caption{Gaussian fit}
        \label{fig:14a}
    \end{subfigure}
    \hfill
    \begin{subfigure}{0.8\textwidth}
        \centering
        \includegraphics[width=\linewidth]{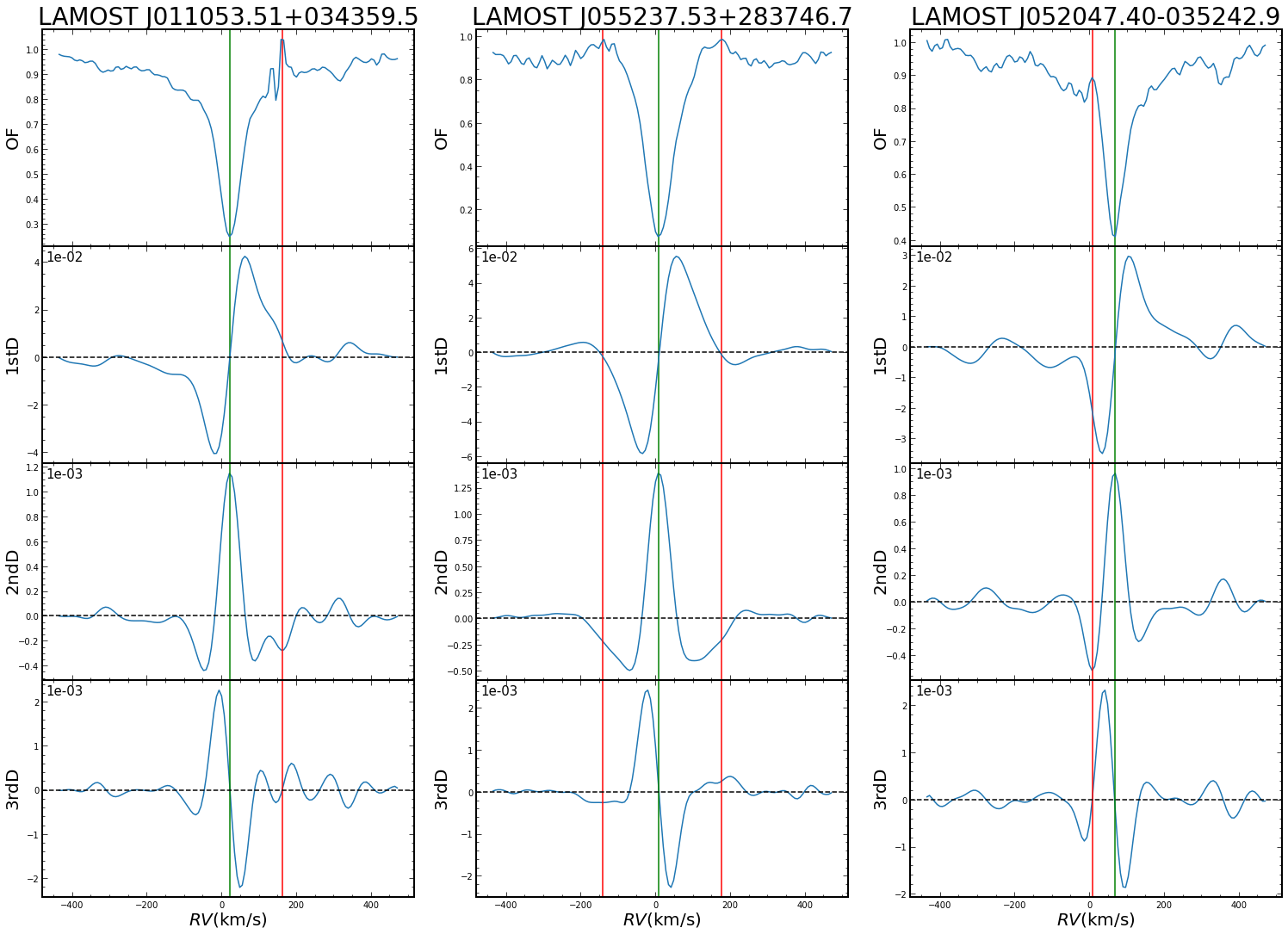}
        \caption{DS}
        \label{fig:14b}
    \end{subfigure}

    \caption{Gaussian fit and DS of J084928.91+103601.5,J043114.44+271017.9,J052847.59+342423.3.}
    \label{Fig14}

\end{figure}

\begin{figure}
    \centering

    \begin{subfigure}{0.8\textwidth}
        \centering
        \includegraphics[width=\linewidth]{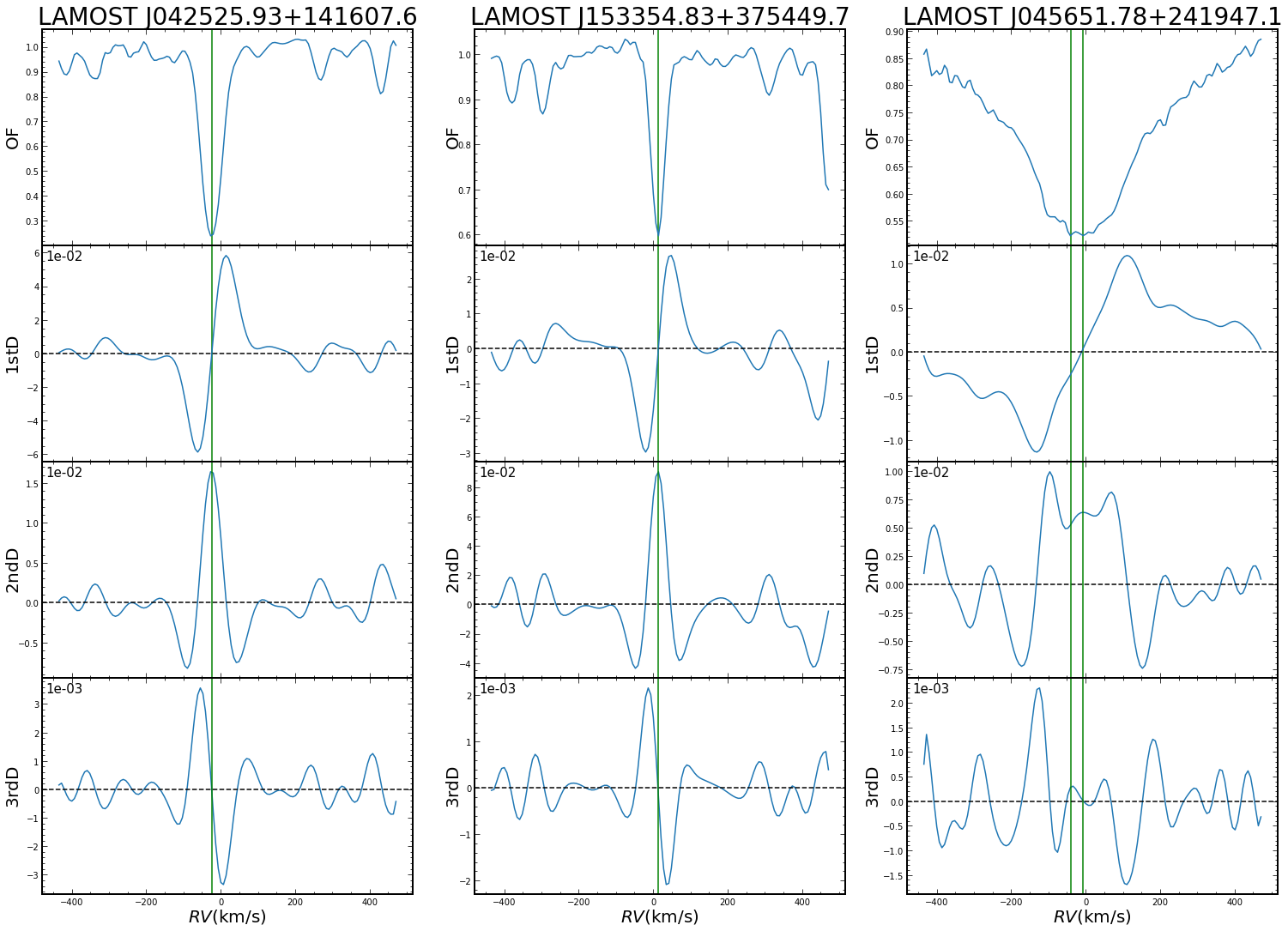}
        \caption{}
        \label{fig:15a}
    \end{subfigure}
    \hfill
    \begin{subfigure}{0.8\textwidth}
        \centering
        \includegraphics[width=\linewidth]{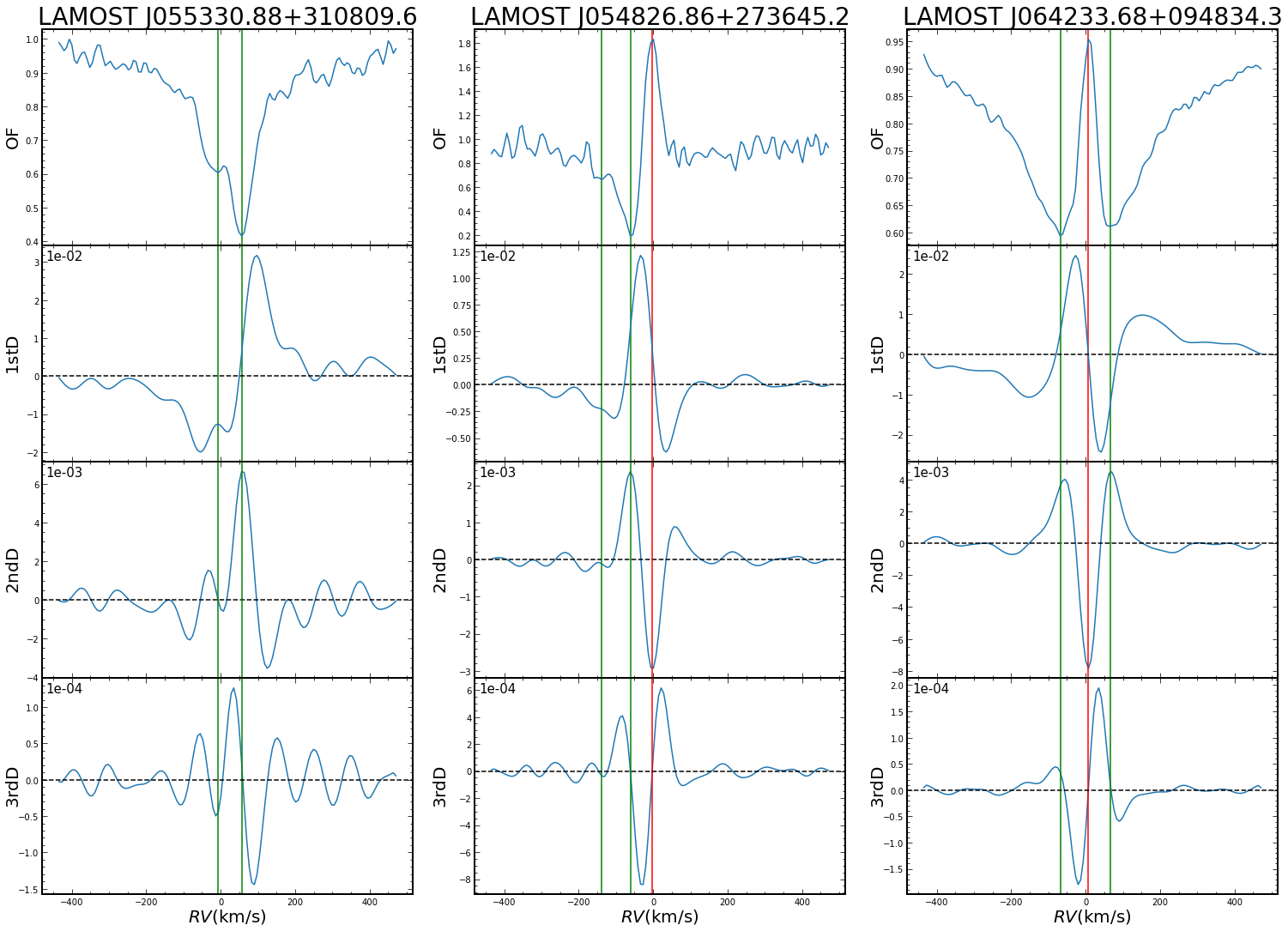}
        \caption{}
        \label{fig:15b}
    \end{subfigure}

    \caption{(a) Examples identified as emission lines by Machine Learning but missed by DS.(b) Examples identified as emission lines by DS but missed by Machine Learning.}
    \label{Fig15}
\end{figure}

This advantage also applies to distinguishing spectra in LAMOST-MRS. We compared the two methods in spectra with confirmed emission lines, and the results are illustrated in Fig. \ref{Fig13} and \ref{Fig14}. It's obvious that the DS method could detect each peak of six spectra of either Fig \ref{Fig13} or Fig \ref{Fig14}. As seen in Fig \ref{Fig13}, two Gaussian functions can be easily fitted to the emission and absorption peaks of the spectrum. For the spectrum of J043114.44+271017.9, which has only one emission peak, both Gaussian functions are fitted to the emission peak, resulting in a fit that closely matches the actual spectrum. However, in Fig \ref{Fig14}, the Gaussian fitting method is more likely to fit two Gaussian functions to the absorption lines with larger amplitudes and broader widths, while overlooking weak and low-resolution emission peak signals. As for the spectrum of J043114.44+271017.9, which has two emission peaks and one absorption peak, or even more complex spectral structures, it becomes challenging for two Gaussian functions to perfectly represent its structural features. Additional Gaussian functions are needed to match different components. This not only increases the difficulty of fitting but also makes it prone to overfitting. In large-scale spectroscopic data from sky survey telescopes, it is challenging to make an initial judgment on the profile of each spectrum and use a different number of Gaussian functions for fitting based on its characteristics.

We compared the DS method, using the parameters outlined in Table \ref{Tab1}, with Machine Learning on both LAMOST-LRS and LAMOST-MRS spectra. In the catalog of \cite{Zhang+2022}, they listed 30,048 spectra linked to 25,886 stars using machine learning. In the comparison, we identified 30,098 spectra from 24,199 stars as emission lines, all of these stars align with \cite{Zhang+2022} findings. The discrepancy in the number of spectra might arise from repeated observations of certain stars. Despite minor differences between the methods, our approach demonstrates higher efficiency as it doesn't necessitate manual re-inspection of spectra. Additionally, the DS method shows greater stability in detecting repeated observations of individual stars.

We also attempted to replicate the machine learning method by calculating equivalent widths or full-width at half-maximum (FWHM) to distinguish emission lines in LAMOST-MRS spectra. We divided 4375 samples into training and testing sets. Then, we tested two machine learning methods, that performed the best in low-resolution spectra, K-Nearest Neighbors (KNN) and Random Forest (RF)\citep{Zhang+2022}. The accuracy in the testing sets of KNN and RF are 0.878 and 0.907 respectively, which is much lower than the 0.997 and 0.989, obtained by \cite{Zhang+2022} at LAMOST-LRS. This may be due to the selection effect and the higher resolution of medium-resolution spectra. In \cite{Zhang+2022}, emission lines were only detected in O, B, and A type, which may to some extent enhance the accuracy. In comparison to LAMOST-LRS, LAMOST-MRS exhibits more morphological features in the $\rm{H\alpha}$ profile. This makes it difficult for features like equivalent width to describe the overall shape of the profile, resulting in reduced accuracy.

We randomly selected 1000 spectra from the catalog provided by DF, which were potentially equipped with emission lines, and another 1000 spectra detected by DS as lacking emission lines. We used these samples to compare the two methods. Machine learning authenticated 1048 spectra as having emission lines. Among them, 157 spectra were authenticated by machine learning as having emission lines, but DF identified them as lacking emission lines. Additionally, 109 spectra were identified by DF as having emission lines, but machine learning classified them as lacking emission lines. We selected a few spectra from each category and plotted them in Fig \ref{Fig15}. The $\rm{H\alpha}$ profiles identified as emission lines by Machine Learning but missed by DS are shown in Fig \ref{fig:15a}, which resemble single absorption lines rather than emission lines. The other situation is shown in Fig \ref{fig:15b}, which has at least one emission peak.

Machine learning outperforms in tasks related to data mining and binary classification. Due to this, DS didn't exhibit a significant edge in detecting faint emission lines. This might be attributed to the relative immaturity of the classification criteria we presently employ. Nevertheless, relying solely on parameters like equivalent width and FWHM for emission line detection, the machine learning approach evidently falls short in estimating the spectral profile and carrying out morphological classification as proficiently as DS and Gaussian fitting. In instances where the spectral profile is intricate, it encounters a challenge akin to Gaussian fitting. This challenge arises from the fact that depending solely on a limited set of characteristic parameters, such as equivalent width and FWHM, makes it arduous to characterize the structural attributes of the spectrum line comprehensively. Ultimately, this leads to a certain degree of misjudgment. Herein lies the strength of derivative spectroscopy. For spectra with profiles of any shape, we can readily differentiate them and identify the parameters linked to their peaks across various orders of derivatives. This empowers us to evaluate their spectral line structure.

\section{Conclusion}
\label{sec:con}

Derivative spectroscopy offers a potential avenue for the morphological analysis of faint signal profiles in spectra. At the testing in LAMOST-MRS, from the 579680 coadded spectra in LAMSOT-MRS, we found 16629 spectra with some kind of features that might be related to the underlying physics of the observed object. The wavelength and amplitude of peaks obtained from the first and third derivatives of the spectrum enable us to construct a simplified morphological classification. We use the classification scheme described in Table \ref{Tab3} to classify all spectra into nine sub-classes.

In comparison to Gaussian fitting, this method exhibits higher sensitivity in detecting faint signals. It enables precise localization of peaks even with small amplitudes and low resolutions, offering accurate wavelength positions and amplitudes. Unlike machine learning, this method provides a more intricate and precise detection process. It can offer an initial assessment of the spectral profile based on the identified peak wavelengths and amplitudes. In fact, DS can furnish seven parameters pertaining to the spectrum, utilizing them to establish a preliminary estimation of the spectral profile. Whether employing this estimation as a prior for spectral fitting or utilizing the current parameters as features and subsequently applying machine learning to refine our existing classification criteria, both approaches are viable. However, this subsequent work is not within the scope of this paper but may be pursued in future studies.

\section*{Acknowledgement}

We would like to express our gratitude for the support provided by the National Natural Science Foundation of China, under grant numbers 12090040/3, 12125303, 12288102, and 11733008. Additionally, we acknowledge the assistance of the National Key Research and Development Program of China (Grant No. 2021YFA1600401/3), the China Manned Space Project (CMS-CSST-2021-A10), and the Yunnan Fundamental Research Projects (Grant No. 202101AV070001). We also extend our appreciation for the joint funding from the National Natural Science Foundation of China and the Chinese Academy of Sciences, under grant No. U1831125, as well as the Research Program of Frontier Sciences, CAS (Grant No. QYZDY-SSW-SLH007). We are grateful for the computational resources provided by the Yunnan Observatories through the Phoenix Supercomputing Platform. Finally, we acknowledge the essential contribution of the Guoshoujing Telescope (LAMOST), a National Major Scientific Project initiated and funded by the Chinese Academy of Sciences, and operated and managed by the National Astronomical Observatories, Chinese Academy of Sciences.

\bibliographystyle{raa}
\bibliography{bibtex}

\end{document}